\documentclass[onecolumn,12pt,journal,draftclsnofoot]{IEEEtran}
\usepackage[latin9]{inputenc}
\usepackage{geometry}
\usepackage{float}
\usepackage{multirow}
\usepackage{amsmath}
\usepackage{amssymb}
\usepackage{graphicx,color}

\makeatletter

\floatstyle{ruled}
\newfloat{algorithm}{tbp}{loa}
\providecommand{\algorithmname}{Algorithm}
\floatname{algorithm}{\protect\algorithmname}

\usepackage{amsfonts}
\usepackage{mathrsfs}
\usepackage{mathrsfs}\usepackage[noblocks]{authblk}
\usepackage[compress]{cite}
\usepackage{bm}
\usepackage{algorithm}
\usepackage{algorithmic}
\usepackage{epsfig}\usepackage{graphics}\usepackage{subfigure}\usepackage{epsfig}\usepackage{epstopdf}
\usepackage{graphics}\usepackage{subfigure}\usepackage{upgreek}\usepackage{theorem}
\usepackage[font=small]{caption}
\usepackage{breqn}
\usepackage{multicol}
\usepackage{eufrak}
\usepackage{eucal}
\usepackage{array}
\usepackage[compress]{cite}
\usepackage{algorithm}
\usepackage{algorithmic}

\allowdisplaybreaks[4]

\newtheorem{theorem}{Theorem}\newtheorem{lemma}{Lemma}\theoremheaderfont{\normalfont\bfseries}

\makeatother

\begin{document}
\title{A Framework for Transmission Design for Active RIS-Aided Communication with Partial CSI}
\author{Gui Zhou, Cunhua Pan, Hong Ren, Dongfang Xu, Zaichen Zhang,   Jiangzhou Wang,  \emph{IEEE Fellow}, and Robert Schober,  \emph{IEEE Fellow}
 \thanks{(Corresponding author: Cunhua Pan)  Part of this work has been published in WCSP Conference \cite{gui-WCSP}.

G. Zhou, D. Xu, and R. Schober are with the Institute for Digital Communications, Friedrich-Alexander-University Erlangen-N\"{u}rnberg (FAU), 91054 Erlangen, Germany (email: gui.zhou, dongfang.xu, robert.schober@fau.de).
C. Pan, H. Ren, and Z. Zhang are with the National Mobile
Communications Research Laboratory, Southeast University, Nanjing
210096, China. (cpan, hren, zczhang@seu.edu.cn). 
J. Wang is with the School of Engineering, University of Kent, Canterbury CT2 7NT, U.K. (e-mail: j.z.wang@kent.ac.uk). }}
\maketitle
\begin{abstract}
Active reconfigurable intelligent surfaces (RISs) have recently been
proposed to compensate for the severe multiplicative fading
effect of conventional passive RIS-aided systems. Each reflecting
element of active RISs is assisted by an amplifier such that the incident
signal can be reflected and amplified instead of only being reflected
as in passive RIS-aided systems.  
This work addresses the practical challenge that, on the one hand, in active RIS-aided systems the  perfect individual CSI of the RIS-aided channels cannot be acquired due to the lack of signal processing power at the  active RISs, but, on the other hand,   this  CSI is required to calculate the expected system data rate and RIS transmit power needed for  transceiver design.
To address this issue, we first derive closed-form expressions for the average achievable rate and the average RIS transmit power based on partial CSI of the RIS-aided channels. Then, we formulate an average achievable rate maximization problem for jointly optimizing the active beamforming at both the base station (BS) and the RIS. This problem is then tackled using the majorization--minimization (MM) algorithm framework, and, for each iteration, semi-closed-form solutions for the BS and RIS beamforming are derived based on 
the Karush-Kuhn-Tucker (KKT) conditions. To ensure the quality of service (QoS)  of each user, we further formulate a rate outage constrained beamforming
problem,   which is solved using the Bernstein-Type inequality (BTI)
and semidefinite relaxation (SDR) techniques. Numerical results show
that the proposed algorithms can efficiently overcome the challenges imposed by imperfect CSI in active RIS-aided wireless systems.
\end{abstract}

\begin{IEEEkeywords}
Reconfigurable intelligent surface (RIS), intelligent reflecting surface
(IRS), active RIS, beamforming, partial channel state information
(CSI).
\end{IEEEkeywords}

\section{Introduction}

Reconfigurable intelligent surfaces (RISs) have attracted extensive research
attention from both academia and industry thanks to their  appealing
features of low cost, low power consumption, programmability, and easy
deployment \cite{qingqing2020mag,cunhua2021mag}. In fact, they are envisioned to be one of the key candidate technologies of sixth generation (6G) mobile 
communication systems \cite{xiaohu2021SINCE,saad2020network}. The existing literature has mainly focused on the investigation of passive RISs, where each reflecting element can only reflect the incident signals. However, passive RISs have  an inherent disadvantage: the   signals reflected by the RISs suffer from multiplicative fading, which  causes the received  signal  to be  extremely weak. Multiplicative fading implies that the equivalent pathloss of the transmitter-RIS-receiver link is the product of the transmitter-RIS link pathloss and the RIS-receiver link pathloss, which is typically thousands of times higher than that of the unobstructed  direct BS-receiver link  \cite{RISmodel2021}. Most of the existing works on passive RISs bypass this issue by assuming a much larger pathloss exponent for  the direct link than for the reflected  links \cite{Pan2019intelleget,Shen2019secrecy}.

To overcome the multiplicative fading effect, the authors of \cite{liangchang2021}
and \cite{daiactiveris2021} recently proposed a new active RIS architecture.
Unlike passive RISs, active RISs are additionally equipped with active
reflective amplifiers. Therefore, active RISs can not only adjust the phase
of the reflected signal, but also amplify the reflected signal. The authors of \cite{daiactiveris2021} showed that in an application scenario with  direct links, passive RISs can only obtain a 3\% data rate gain,
while active RISs can obtain a 108\% gain. In addition, they 
 also presented a hardware platform for  active RISs.

Different from traditional active antenna arrays, active RISs do not require radio frequency (RF) chains and digital signal processing circuits, such that active RISs can be relatively thin,  which  facilitates deployment. 
Since active RISs comprise  amplifiers, their hardware power consumption is increased compared to passive RISs. However, the authors of \cite{zhi2021active} recently compared the performances of passive RISs and  active RISs for  the same total power consumption (including hardware power consumption), and showed that active RISs outperform passive RISs when the number of reflecting elements is small and the
system power budget is sufficiently large. Therefore, active RISs can mitigate the multiplicative fading effect  while retaining the benefits  of passive RISs.

Due to  the above advantages, active RISs have attracted significant
research interest recently. The authors of \cite{changsheng2021active}
compared the performances of passive RISs and active RISs, and optimized the RIS location and number of reflecting elements. The authors of  \cite{dongfang2021active} investigated the resource allocation design  for active RIS-aided multiuser systems. Furthermore, active RISs have  been considered for  wireless powered communications to  enhance throughput and energy efficiency \cite{qingqingacttive}.

It is widely known that, for passive RIS-aided systems, only the cascaded channel state information (CSI) of the transmitter-RIS-receiver link is needed for transceiver design. However, due to the introduction of amplifiers,  for active RIS-aided systems,  the  RIS transmit power and the thermal noise amplified by the RISs need to be taken into account  for transceiver design, which requires the individual CSI of the transmitter-RIS link and the RIS-receiver link. However, active RISs cannot estimate the two individual channels as they are not equipped  with an RF chain. To the best of the authors'  knowledge, all existing works  on active RISs assume the availability of perfect   CSI of the transmitter-RIS  and RIS-receiver links, respectively, which is challenging to obtain in practice. Therefore, it is imperative to study the system design for the case,  where only partial  CSI of the individual active RIS-aided channels is available.

Against this  background, in this work,  average achievable rate maximization and average power consumption minimization are addressed, respectively, if only partial  CSI of the individual RIS-aided channels is available.  To this end, we assume that the RIS-aided channels are Rician distributed. Although the perfect CSI of the individual RIS-aided channels is not available,  knowledge of the deterministic light-of-sight (LoS) components  and the statistics of the Gaussian distribution of the non-LoS (NLoS) components can be acquired. In particular, the angle and distance information of the LoS links can be determined via localization techniques \cite{panJSTSP2022}. Based on this partial CSI, we derive analytical expressions for the  average achievable rate and the average RIS transmit power. Then, the average achievable rate is maximized by jointly optimizing the active beamforming at the BS and RIS under an RIS average transmit power constraint. Since system designs based on average achievable rate maximization cannot guarantee the QoS of each user, we further study designs based on a  rate outage constrained power minimization problem.

The main contributions of this work can be summarized as follows:
\begin{itemize}
\item To the best of the authors'  knowledge, this is the first work on active
RIS-aided systems that investigates the practical issue of partial CSI knowledge.
Based on the distributions of the individual RIS-aided channels, we propose a
 joint active beamforming design at the BS and the RIS for maximization of the average achievable rate for partial CSI. In addition, we also study the  robust
active beamforming design to minimize the average total power consumption under rate outage probability constraints.
\item Closed-form expressions for  the average achievable rate and the average RIS transmit
power in the presence of partial CSI are  derived.  Furthermore, the average achievable rate maximization
problem is efficiently solved in an iterative manner exploiting  the majorization--minimization (MM) concept. Specifically, a surrogate quadratic function for  active
beamforming is  constructed to minorize the original non-concave
objective function. Then, alternating optimization (AO) is employed 
to decouple the BS and RIS beamforming vectors. For
each subproblem, a semi-closed-form solution is obtained based on  the Karush--Kuhn--Tucker (KKT) conditions.
\item To guarantee a predefined outage probability, we develop an outage
constrained beamforming design that minimizes the average transmit
power subject to constraints on the RIS amplification gain and the rate outage probability, respectively.  The Bernstein-type inequality (BTI) is applied
to safely approximate the  outage probability constraints such
that the non-convexity of the constraints is mitigated. Then, the beamforming
vectors at both the BS and the RIS are updated by using semidefinite
relaxation (SDR) in an iterative manner.
\item Our simulation results demonstrate that active RISs can effectively overcome the negative impact of the multiplicative fading effect and perform much better than the conventional passive RISs.  Furthermore, since the RIS amplifier circuits  consume power, there exists an optimal number of RIS
reflecting elements.
\end{itemize}
The rest of this paper is organized as follows. In Section II, we  introduce
the considered system model. The average achievable rate maximization problem and the average
power minimization problem are respectively revealed in Sections III
and IV. Finally, Sections V and VI report numerical results
and conclusions, respectively.

\textbf{Notations:} The following mathematical notations and symbols
are used throughout this paper. Vectors and matrices are denoted by
boldface lowercase letters and boldface uppercase letters, respectively.
 $\mathbf{X}^{*}$, $\mathbf{X}^{\mathrm{T}}$, $\mathbf{X}^{\mathrm{H}}$,
and $||\mathbf{X}||_{F}$ denote the conjugate, transpose, Hermitian
(conjugate transpose), and Frobenius norm of matrix $\mathbf{X}$, respectively.
$\mathrm{vec}(\mathbf{X})$ denotes the vectorization of matrix $\mathbf{X}$.
$||\mathbf{x}||_{2}$ denotes the L2-norm of vector $\mathbf{x}$.
Operations $\mathrm{Tr}\{\cdot\}$, $\mathrm{Re}\{\cdot\}$, $|\cdot|$,
$\lambda(\cdot)$, and $\angle\left(\cdot\right)$ denote the trace,
real part, modulus, eigenvalue, and angle of a complex number, respectively.
$\mathrm{Diag}(\mathbf{x})$ is a diagonal matrix with the entries
of $\mathbf{x}$ on its main diagonal. Furthermore, $\mathrm{diag}(\mathbf{X})$
is a vector whose entries are the main diagonal elements of matrix $\mathbf{X}$.
$[\mathbf{x}]_{m}$ denotes the $m$-th element of vector $\mathbf{x}$. $[\mathbf{X}]_{m:n,p:q}$ is a matrix consisting of the $m$-th to the $n$-th rows and the  $p$-th  to the $q$-th columns  of matrix $\mathbf{X}$.
The Kronecker product, Hadamard product, and Khatri-Rao product between
two matrices $\mathbf{X}$ and $\mathbf{Y}$ are respectively denoted
by $\mathbf{X}\otimes\mathbf{Y}$, $\mathbf{X}\odot\mathbf{Y}$, and
$\mathbf{X}\diamond\mathbf{Y}$. $\mathbf{X}\succeq\mathbf{Y}$ means
that $\mathbf{X}-\mathbf{Y}$ is positive semidefinite. 
$\mathbb{C}$ denotes the complex field, $\mathbb{R}$ denotes the
real field, and $j\triangleq\sqrt{-1}$ is the imaginary unit. $\mathcal{CN}(\boldsymbol{\mu},\boldsymbol{\Sigma})$ {represents} the distribution of a circularly symmetric complex Gaussian random vector with mean vector $\boldsymbol{\mu}$ and  covariance matrix $\boldsymbol{\Sigma}$. 

\section{System Model}

\subsection{Signal Transmission Model}

\begin{figure}
\centering \includegraphics[width=3.6in,height=2.9in]{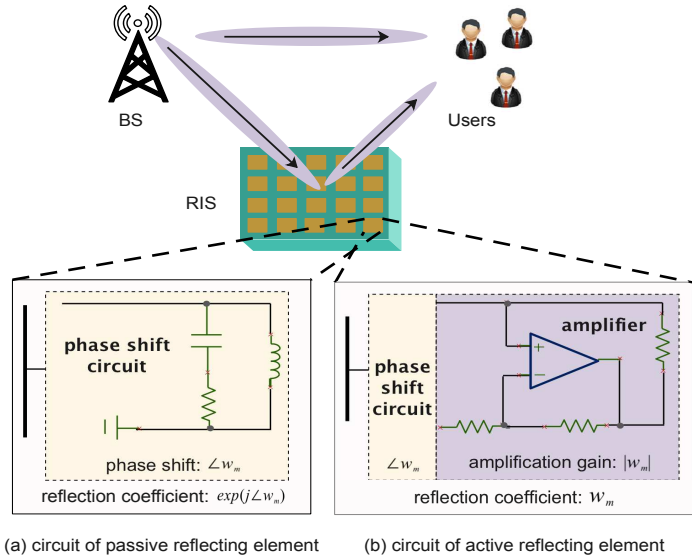}
\caption{Active and passive RIS-aided communication system, respectively.}
\label{system-model}
\end{figure}

As shown in Fig. \ref{system-model}, we consider an RIS-aided downlink multiple-input single-output (MISO) system, where an $N$-antenna BS communicates with $K$ single-antenna users. The RIS is assumed to be equipped with $M$ reflecting elements, and its reflection coefficient matrix is given by  $\boldsymbol{\Lambda}_{\mathbf{w}}=\mathrm{Diag}(w_{1},\cdots,w_{M})\in\mathbb{C}^{M\times M}$. Here, $\angle w_{m}$ and $|w_{m}|$ denote the phase shift and the reflection gain of the $m$-th RIS element, respectively. For passive RISs, each RIS element comprises an impedance adjustable circuit to vary the phase shift \cite{shanpu2022}. Thus,  passive RISs are capable of reflecting the incident signal\footnote{Here, we assume an ideal reflective material without reflection loss. If reflection loss is considered, then $|w_{m}|^{2}<1$.} without consuming direct-current (DC) power, which leads to a reflection gain of $|w_{m}|^{2}=1$ and negligible thermal noise. However, the multiplicative fading effect results in a weak received signal power for the passive RIS reflection link. To address this issue, the authors of \cite{liangchang2021} and \cite{daiactiveris2021} proposed a new active RIS architecture, where each active RIS element includes an  active reflection-type amplifier to also amplify the incident signals. Therefore, the reflection gain is given by $1\leq|w_{m}|^{2}\leq a_{max}$, where $a_{max}$ is the maximum amplification gain.

The BS transmits $K$ data symbols collected in vector $\mathbf{s}=[s_{1},\cdots,s_{K}]^{\mathrm{T}}\in\mathbb{C}^{K\times1}$
to the users by applying precoder matrix $\mathbf{F=}[{\bf \mathbf{f}}_{1},\cdots,{\bf \mathbf{f}}_{K}]\in\mathbb{C}^{N\times K}$.
By assuming independent complex Gaussian signals with  $\mathbb{E}[\mathbf{s}\mathbf{s}^{\mathrm{H}}]=\mathbf{I}_{N}$,
the BS transmit power is given by $\mathbb{E}\{||{\bf F}{\bf s}||_{2}^{2}\}=||\mathbf{F}||_{F}^{2}\leq P_{N}$,
where $P_{N}$ is the BS transmit power budget. Denote by
$\mathbf{H_{\mathrm{dr}}}\in\mathbb{C}^{M\times N}$ the channel from
the BS to the RIS, and by $\mathbf{h}_{k}\in\mathbb{C}^{N\times1}$ and
$\mathbf{h}_{\mathrm{r},k}\in\mathbb{C}^{M\times1}$ the channels
from user $k$ to the BS and to the RIS, respectively. Then, the signal
received by user $k$ is given by
\begin{align}
y_{k} & =\mathbf{h}_{k}^{\mathrm{H}}{\bf F}{\bf s}+\mathbf{h}_{\mathrm{r},k}^{\mathrm{H}}\boldsymbol{\Lambda}_{\mathbf{w}}(\mathbf{H_{\mathrm{dr}}}{\bf F}{\bf s}+{\bf z})+n_{k}\nonumber \\
 & =(\mathbf{h}_{k}^{\mathrm{H}}+\mathbf{h}_{\mathrm{r},k}^{\mathrm{H}}\boldsymbol{\Lambda}_{\mathbf{w}}\mathbf{H_{\mathrm{dr}}}){\bf F}{\bf s}+\mathbf{h}_{\mathrm{r},k}^{\mathrm{H}}\boldsymbol{\Lambda}_{\mathbf{w}}{\bf z}+n_{k},
\end{align}
where $n_{k}$ and ${\bf z}$ are the zero-mean additive white Gaussian noise
(AWGN) at the user and the RIS, respectively, which follow  distributions
$n_{k}\sim\mathcal{CN}(0,\sigma_{k}^{2})$ and ${\bf z}\sim\mathcal{CN}(\mathbf{0},\sigma_{z}^{2}{\bf I}_{M})$ with noise powers  $\sigma_{k}^{2}$ and $\sigma_{z}^{2}$, respectively. Notice that the thermal noise ${\bf z}$, which can be ignored
in passive RIS-aided systems, has to be considered in active RIS-aided
systems because of  the amplification. The  transmit power of the active RIS is given by
\begin{equation}
\mathbb{E}\{||\boldsymbol{\Lambda}_{\mathbf{w}}(\mathbf{H_{\mathrm{dr}}}{\bf F}{\bf s}+{\bf z})||_{2}^{2}\}=||\boldsymbol{\Lambda}_{\mathbf{w}}\mathbf{H_{\mathrm{dr}}}{\bf F}||_{F}^{2}+||\mathbf{w}||_{2}^{2}\sigma_{z}^{2},\label{eq:rispower}
\end{equation}
where $\mathbf{w}=[w_{1},\cdots,w_{M}]^{\mathrm{H}}$. {Furthermore, the} achievable rate of user $k$ is given by 
\begin{align}
R_{k}({\bf F},{\bf w})=\log_{2}\left(1+\frac{|(\mathbf{h}_{k}^{\mathrm{H}}+\mathbf{h}_{\mathrm{r},k}^{\mathrm{H}}\boldsymbol{\Lambda}_{\mathbf{w}}\mathbf{H_{\mathrm{dr}}}){\bf f}_{k}|^{2}}{\sum_{i=1,i\neq k}^{K}|(\mathbf{h}_{k}^{\mathrm{H}}+\mathbf{h}_{\mathrm{r},k}^{\mathrm{H}}\boldsymbol{\Lambda}_{\mathbf{w}}\mathbf{H_{\mathrm{dr}}}){\bf f}_{i}|^{2}+\sigma_{k}^{2}+||\mathbf{h}_{\mathrm{r},k}^{\mathrm{H}}\boldsymbol{\Lambda}_{\mathbf{w}}||_{2}^{2}\sigma_{z}^{2}}\right).\label{eq:rate}
\end{align}

\subsection{Channel Model}

In practice, not all of the channels connecting the BS and the users
can be individually estimated. Specifically, the direct BS-user channels
$\{\mathbf{h}_{k}\}_{k=1}^{K}$ can be estimated by turning off the
RIS \cite{shuguang-IRS}. Thus, it is reasonable to assume that perfect
CSI of the direct BS-user channels is available at the BS. However, for
the passive/active RIS-aided channels, we can only estimate the
cascaded BS-RIS-user channel $\mathbf{G}_{k}$, which is the product
of the BS-RIS channel $\mathbf{H_{\mathrm{dr}}}$ and the RIS-user
channel $\mathbf{h}_{\mathrm{r},k}$, i.e., $\mathbf{G}_{k}=\mathrm{diag}(\mathbf{h}_{\mathrm{r},k}^{\mathrm{H}})\mathbf{H_{\mathrm{dr}}}$.
As a result, we cannot estimate $\mathbf{H_{\mathrm{dr}}}$ and $\mathbf{h}_{\mathrm{r},k}$
individually due to the lack of signal processing capability at the
passive/active RIS  \cite{Lee-proceeding}. There is an extensive literature on cascaded
CSI estimation in RIS-aided communication systems \cite{shuguang-IRS,weiyi2022twc,ris-omp-3,gui2022channel}.
Thus, in this work, we also assume that $\{\mathbf{G}_{k}\}_{k=1}^{K}$ is perfectly known
at the BS. In  passive RIS-aided communication systems, knowledge of the CSI
of the cascaded channels is typically sufficient for  transceiver and  RIS reflection
phase shift design \cite{Gui2019IRS,Xianghao2009,Shen2019secrecy,OFDM2019}.
However, in active RIS-aided systems, the instantaneous RIS transmit power
in (\ref{eq:rispower}) and the instantaneous achievable rate in (\ref{eq:rate}) depend on the individual instantaneous CSI of $\mathbf{H_{\mathrm{dr}}}$
and $\mathbf{h}_{\mathrm{r},k}$, respectively, and this CSI is impossible
to obtain. To address this issue, in this work, we focus on the investigation
of the average achievable rate and the average RIS transmit power based on statistical CSI of $\mathbf{H_{\mathrm{dr}}}$ and $\mathbf{h}_{\mathrm{r},k}$.

In particular, $\mathbf{H_{\mathrm{dr}}}$ and $\mathbf{h}_{\mathrm{r},k}$
are modelled as correlated Rician fading channels as follows
\begin{align}
{\mathbf{H}}_{\mathrm{dr}} & =\sqrt{\beta_{0}/(\delta_{0}+1)}(\sqrt{\delta_{0}}\bar{\mathbf{H}}_{\mathrm{dr}}+\tilde{\mathbf{H}}_{\mathrm{dr}}),\label{eq:re}\\
\mathbf{h}_{\mathrm{r},k} & =\sqrt{\beta_{k}/(\delta_{k}+1)}(\sqrt{\delta_{k}}\bar{\mathbf{h}}_{\mathrm{r},k}+\tilde{\mathbf{h}}_{\mathrm{r},k}),\forall k,\label{eq:tr}
\end{align}
where $\{\delta_{k}\}_{k=0}^{K}$ are the Rician factors, and  $\{\beta_{k}\}_{k=0}^{K}$
are the distance-dependent large-scale pathloss coefficients. The
LoS components $\bar{\mathbf{H}}_{\mathrm{dr}}$ and $\{\bar{\mathbf{h}}_{\mathrm{r},k}\}_{k=1}^{K}$
are determined by the angles-of-arrival (AoAs) and the angles-of-departure
(AoDs) \cite{book-Goldsmith}. The physical positions of the BS and the RIS are generally
fixed and known in advance, and the users' locations can be determined
by GPS positioning \cite{abu2018error} or pilot-based positioning
algorithms \cite{Elzanaty2021position}. Thus, the communication distance
and LoS angle information can be  assumed to be known by
the BS. The NLoS components are distributed as  $\tilde{\mathbf{H}}_{\mathrm{dr}}\sim\mathcal{CN}(\mathbf{0},\boldsymbol{\Sigma}_{\mathrm{R}}\otimes\boldsymbol{\Sigma}_{\mathrm{B}})$
and $\tilde{\mathbf{h}}_{\mathrm{r},k}\sim\mathcal{CN}(\mathbf{0},\boldsymbol{\Sigma}_{\mathrm{r},k})$,
where $\boldsymbol{\Sigma}_{\mathrm{B}}\succeq\mathbf{0}$ is the
spatial correlation matrix with unit diagonal elements at the BS for
channel $\tilde{\mathbf{H}}_{\mathrm{dr}}$, and $\boldsymbol{\Sigma}_{\mathrm{R}}\succeq\mathbf{0}$
and $\boldsymbol{\Sigma}_{\mathrm{r},k}\succeq\mathbf{0}$ are the
spatial correlation matrices with unit diagonal elements at the RIS
for channels $\tilde{\mathbf{H}}_{\mathrm{dr}}$ and $\tilde{\mathbf{h}}_{\mathrm{r},k}$,
respectively. The spatial covariance matrices can be estimated with the method proposed in \cite{changliang} or the model proposed in \cite{Bjornson2021}. Thus, in the following, we model  $\tilde{\mathbf{H}}_{\mathrm{dr}}$ and $\tilde{\mathbf{h}}_{\mathrm{r},k}$
 as $\tilde{\mathbf{H}}_{\mathrm{dr}}=\boldsymbol{\Sigma}_{\mathrm{R}}^{1/2}\mathbf{E}\boldsymbol{\Sigma}_{\mathrm{B}}^{1/2}$
with $\mathrm{vec}(\mathbf{E})\sim\mathcal{CN}(\mathbf{0},\mathbf{I}_{M}\otimes\mathbf{I}_{N})$
and $\tilde{\mathbf{h}}_{\mathrm{r},k}=\boldsymbol{\Sigma}_{\mathrm{r},k}^{1/2}\mathbf{e}_{\mathrm{r},k}$
with $\mathbf{e}_{\mathrm{r},k}\sim\mathcal{CN}(\mathbf{0},\mathbf{I}_{M})$,
respectively.

Since the communication distances between the BS and the users are
generally long and the electromagnetic environment is complex, we assume  channels $\{\mathbf{h}_{k}\}_{k=1}^{K}$ to be Rayleigh distributed.

\subsection{Average Achievable Rate and Average RIS Transmit Power}
Since perfect instantaneous CSI of $\mathbf{H_{\mathrm{dr}}}$ and $\{\mathbf{h}_{\mathrm{r},k}\}_{k=1}^{K}$ is not available, in this work, we  consider the average achievable rate and average RIS transmit power,  denoted as $\mathbb{E}_{\mathbf{h}_{\mathrm{r},k}|{\bf G}_k}\left\{ R_{k}({\bf F},{\bf w})\right\}$   for all $k$ and $\mathbb{E}_{\mathbf{H_{\mathrm{dr}}}|{\bf G}_k}\left\{ ||\boldsymbol{\Lambda}_{\mathbf{w}}\mathbf{H_{\mathrm{dr}}}{\bf F}||_{F}^{2}+||\mathbf{w}||_{F}^{2}\sigma_{z}^{2}\right\}$, respectively.  Here, the average rate and average RIS transmit power are short-term (instantaneous) measures that capture the expected performance over the distributions of $\{\mathbf{h}_{\mathrm{r},k}\}_{k=1}^K$ and $\mathbf{H_{\mathrm{dr}}}$ for given $\{\mathbf{G}_{k}\}_{k=1}^{K}$.

First, we derive an {analytical} expression for the average achievable rate. With the definitions $\tilde{\mathbf{w}} = [ 	\mathbf{w}^{\mathrm{H}}, 1 ]^{\mathrm{H}}$ and $\mathbf{H}_{k} = \left[
\mathbf{G}_{k}^{\mathrm{H}} ,  	\mathbf{h}_{k}\right]^{\mathrm{H}}$ such that   $\tilde{\mathbf{w}}^{\mathrm{H}}\mathbf{H}_{k} = \mathbf{h}_{k}^{\mathrm{H}}+\mathbf{h}_{\mathrm{r},k}^{\mathrm{H}}\boldsymbol{\Lambda}_{\mathbf{w}}^{\mathrm{H}}\mathbf{H_{\mathrm{dr}}}$,  (\ref{eq:rate}) can be reformulated as follows
\begin{align}
	R_{k}({\bf F},{\bf w})= & \log_{2}\left(1+\frac{|\tilde{\mathbf{w}}^{\mathrm{H}}\mathbf{H}_{k}{\bf f}_{k}|^{2}}{\sum_{i=1,i\neq k}^{K}|\tilde{\mathbf{w}}^{\mathrm{H}}\mathbf{H}_{k}{\bf f}_{i}|^{2}+\sigma_{z}^{2}||\mathbf{h}_{\mathrm{r},k}^{\mathrm{H}}\boldsymbol{\Lambda}_{\mathbf{w}}^{\mathrm{H}}||_{2}^{2}+\sigma_{k}^{2}}\right).\label{eq:rate-1}
\end{align}
Since the function $f(x)=\log_{2}(1+\frac{1}{x})$ is convex in $x$,
by using Jensen's inequality, we can obtain a lower bound for  $\mathbb{E}_{\mathbf{h}_{\mathrm{r},k}|{\bf G}_k}\left\{ R_{k}({\bf F},{\bf w})\right\} $
as follows
\begin{align}
	&\;\;\;\; \; \mathbb{E}_{\mathbf{h}_{\mathrm{r},k}|{\bf G}_k}\left\{ R_{k}({\bf F},{\bf w})\right\}   \nonumber \\ &\geq\bar{R}_{k}\left(\mathbf{F},\mathbf{w}\right)\nonumber \\
	& =\log_{2}\left(1+\frac{|\tilde{\mathbf{w}}^{\mathrm{H}}\mathbf{H}_{k}{\bf f}_{k}|^{2}}{\sum_{i=1,i\neq k}^{K}|\tilde{\mathbf{w}}^{\mathrm{H}}\mathbf{H}_{k}{\bf f}_{i}|^{2}+\sigma_{z}^{2}\mathbb{E}_{\mathbf{h}_{\mathrm{r},k}|{\bf G}_k}\left\{ ||\mathbf{h}_{\mathrm{r},k}^{\mathrm{H}}\boldsymbol{\Lambda}_{\mathbf{w}}^{\mathrm{H}}||_{2}^{2}\right\} +\sigma_{k}^{2}}\right)\nonumber \\
	& =\log_{2}\left(1+\frac{|\tilde{\mathbf{w}}^{\mathrm{H}}\mathbf{H}_{k}{\bf f}_{k}|^{2}}{\sum_{i=1,i\neq k}^{K}|\tilde{\mathbf{w}}^{\mathrm{H}}\mathbf{H}_{k}{\bf f}_{i}|^{2}+\sigma_{z}^{2}\mathrm{Tr}\left\{ \boldsymbol{\Lambda}_{\mathbf{w}}\mathbb{E}_{\mathbf{h}_{\mathrm{r},k}|{\bf G}_k}\left\{ \mathbf{h}_{\mathrm{r},k}\mathbf{h}_{\mathrm{r},k}^{\mathrm{H}}\right\} \boldsymbol{\Lambda}_{\mathbf{w}}^{\mathrm{H}}\right\} +\sigma_{k}^{2}}\right)\nonumber \\
	& \overset{(a)}{=}\log_{2}\left(1+\frac{|\tilde{\mathbf{w}}^{\mathrm{H}}\mathbf{H}_{k}{\bf f}_{k}|^{2}}{\sum_{i=1,i\neq k}^{K}|\tilde{\mathbf{w}}^{\mathrm{H}}\mathbf{H}_{k}{\bf f}_{i}|^{2}+\sigma_{z}^{2}\mathrm{Tr}\left\{ \boldsymbol{\Lambda}_{\mathbf{w}}\left(\frac{\beta_{k}\delta_{k}}{\delta_{k}+1}\bar{\mathbf{h}}_{\mathrm{r},k}\bar{\mathbf{h}}_{\mathrm{r},k}^{\mathrm{H}}+\frac{\beta_{k}}{\delta_{k}+1}\boldsymbol{\Sigma}_{\mathrm{r},k}\right)\boldsymbol{\Lambda}_{\mathbf{w}}^{\mathrm{H}}\right\} +\sigma_{k}^{2}}\right)\nonumber \\
	& \overset{(b)}{=}\log_{2}\left(1+\frac{|\tilde{\mathbf{w}}^{\mathrm{H}}\mathbf{H}_{k}{\bf f}_{k}|^{2}}{\sum_{i=1,i\neq k}^{K}|\tilde{\mathbf{w}}^{\mathrm{H}}\mathbf{H}_{k}{\bf f}_{i}|^{2}+\sigma_{z}^{2}\left(\frac{\beta_{k}\delta_{k}}{\delta_{k}+1}||\mathrm{Diag}(\bar{\mathbf{h}}_{\mathrm{r},k})\mathbf{w}||_{2}^{2}+\frac{\beta_{k}}{\delta_{k}+1}||\mathbf{w}||_{2}^{2}\right)+\sigma_{k}^{2}}\right)\nonumber \\
	& =\log_{2}\left(1+\frac{|\tilde{\mathbf{w}}^{\mathrm{H}}\mathbf{H}_{k}{\bf f}_{k}|^{2}}{\sum_{i=1,i\neq k}^{K}|\tilde{\mathbf{w}}^{\mathrm{H}}\mathbf{H}_{k}{\bf f}_{i}|^{2}+\mathbf{w}^{\mathrm{H}}\boldsymbol{\Psi}_{k}\mathbf{w}+\sigma_{k}^{2}}\right),\label{eq:rate-a}
\end{align}
where $\boldsymbol{\Psi}_{k}=\frac{\beta_{k}\sigma_{z}^{2}}{\delta_{k}+1}\left(\delta_{k}\mathrm{Diag}(\bar{\mathbf{h}}_{\mathrm{r},k}\odot\bar{\mathbf{h}}_{\mathrm{r},k}^{*})+\mathbf{I}\right)$ and $\mathbf{I}$ denotes the identity matrix.
Equality (a) in (\ref{eq:rate-a}) is obtained due to
$\mathbf{h}_{\mathrm{r},k}\sim\mathcal{CN}\left(\sqrt{\beta_{k}\delta_{k}/(\delta_{k}+1)}\bar{\mathbf{h}}_{\mathrm{r},k},\frac{\beta_{k}}{\delta_{k}+1}\boldsymbol{\Sigma}_{\mathrm{r},k}\right)$
and $\mathbb{E}_{\mathbf{\mathbf{h}_{\mathrm{r},k}}|{\bf G}_k}\left\{ \mathbf{h}_{\mathrm{r},k}\mathbf{h}_{\mathrm{r},k}^{\mathrm{H}}\right\} =\frac{\beta_{k}\delta_{k}}{\delta_{k}+1}\bar{\mathbf{h}}_{\mathrm{r},k}\bar{\mathbf{h}}_{\mathrm{r},k}^{\mathrm{H}}+\frac{\beta_{k}}{\delta_{k}+1}\boldsymbol{\Sigma}_{\mathrm{r},k}$.
Equality (b) in (\ref{eq:rate-a}) is due to $\mathrm{Tr}\left\{ \boldsymbol{\Lambda}_{\mathbf{w}}\bar{\mathbf{h}}_{\mathrm{r},k}\bar{\mathbf{h}}_{\mathrm{r},k}^{\mathrm{H}}\boldsymbol{\Lambda}_{\mathbf{w}}^{\mathrm{H}}\right\} =||\mathrm{Diag}(\bar{\mathbf{h}}_{\mathrm{r},k})\mathbf{w}||_{2}^{2}$
and $\frac{\beta_{k}}{\delta_{k}+1}\mathrm{Tr}\left\{ \boldsymbol{\Lambda}_{\mathbf{w}}\boldsymbol{\Sigma}_{\mathrm{r},k}\boldsymbol{\Lambda}_{\mathbf{w}}^{\mathrm{H}}\right\} \\=\frac{\beta_{k}}{\delta_{k}+1}\mathrm{Tr}\left\{ \boldsymbol{\Lambda}_{\mathbf{w}}\boldsymbol{\Lambda}_{\mathbf{w}}^{\mathrm{H}}\right\} =\frac{\beta_{k}}{\delta_{k}+1}||\mathbf{w}||_{2}^{2}$,
as  $\boldsymbol{\Sigma}_{\mathrm{r},k}$ has unit diagonal elements.

Next, to derive an  analytical expression for the average RIS transmit power, we provide a useful lemma, as follows.

\begin{lemma}\label{lemma-h}
	
	Let $\mathbf{H}\in\mathbb{C}^{M\times N}=\bar{\mathbf{H}}+\boldsymbol{\Sigma}_{\mathrm{r}}^{1/2}\mathbf{H}^{w}\boldsymbol{\Sigma}_{\mathrm{t}}^{1/2}$
	represent a random matrix following distribution  $\mathbf{H}\sim\mathcal{CN}(\bar{\mathbf{H}},\boldsymbol{\Sigma}_{\mathrm{r}}\otimes\boldsymbol{\Sigma}_{\mathrm{t}})$
	with mean $\bar{\mathbf{H}}$ and covariance $\boldsymbol{\Sigma}_{\mathrm{r}}\otimes\boldsymbol{\Sigma}_{\mathrm{t}}$,
	where $\mathbf{H}^{w}$ is a complex Gaussian random matrix with independent and identically distributed (i.i.d.)
	entries of zero mean and unit variance. Given matrix $\mathbf{X}\in\mathbb{C}^{N\times N}$,
	we have
	\[
	\mathbb{E}_{\mathbf{H}}\left\{ \mathbf{H}\mathbf{X}\mathbf{H}^{\mathrm{H}}\right\} =\bar{\mathbf{H}}\mathbf{X}\bar{\mathbf{H}}^{\mathrm{H}}+\mathrm{Tr}\left\{ \mathbf{X}\boldsymbol{\Sigma}_{\mathrm{t}}\right\} \boldsymbol{\Sigma}_{\mathrm{r}}.
	\]
	
\end{lemma}\textbf{\textit{Proof: }}Please refer to the proof of
Lemma 2 in \cite{relay-statictical}.\hspace{8.3cm}$\blacksquare$

By using $\mathbf{H}_{\mathrm{dr}}\sim\mathcal{CN}\left(\sqrt{\frac{\beta_{0}\delta_{0}}{\delta_{0}+1}}\bar{\mathbf{H}}_{\mathrm{dr}},\frac{\beta_{0}}{\delta_{0}+1}\left(\boldsymbol{\Sigma}_{\mathrm{R}}\otimes\boldsymbol{\Sigma}_{\mathrm{B}}\right)\right)$
and Lemma \ref{lemma-h}, the average RIS transmit power $P({\bf F},{\bf w})$ can be obtained as
\begin{align}
	P({\bf F},{\bf w}) & =\mathbb{E}_{\mathbf{H_{\mathrm{dr}}}|{\bf G}_k}\left\{ ||\boldsymbol{\Lambda}_{\mathbf{w}}\mathbf{H}_{\mathrm{dr}}{\bf F}||_{F}^{2}+||\mathbf{w}||_{2}^{2}\sigma_{z}^{2}\right\} \nonumber \\
	& =\mathrm{Tr}\left\{ \boldsymbol{\Lambda}_{\mathbf{w}}\mathbb{E}_{\mathbf{H_{\mathrm{dr}}}}\left\{ \mathbf{H}_{\mathrm{dr}}{\bf F}{\bf F}^{\mathrm{H}}\mathbf{H}_{\mathrm{dr}}^{\mathrm{H}}\right\} \boldsymbol{\Lambda}_{\mathbf{w}}^{\mathrm{H}}\right\} +||\mathbf{w}||_{2}^{2}\sigma_{z}^{2}\nonumber \\
	& =\mathrm{Tr}\left\{ \boldsymbol{\Lambda}_{\mathrm{RIS}}^{a}{\bf Q}\right\} +||\mathbf{w}||_{2}^{2}\sigma_{z}^{2},\label{eq:frfr}
\end{align}
where ${\bf Q}=\left(\frac{\beta_{0}\delta_{0}}{\delta_{0}+1}\bar{\mathbf{H}}_{\mathrm{dr}}{\bf F}{\bf F}^{\mathrm{H}}\bar{\mathbf{H}}_{\mathrm{dr}}^{\mathrm{H}}+\frac{\beta_{0}}{\delta_{0}+1}\mathrm{Tr}\left\{ {\bf F}{\bf F}^{\mathrm{H}}\boldsymbol{\Sigma}_{\mathrm{B}}\right\} \boldsymbol{\Sigma}_{\mathrm{R}}\right)$
and $\boldsymbol{\Lambda}_{\mathrm{RIS}}^{a}=\boldsymbol{\Lambda}_{\mathbf{w}}^{\mathrm{H}}\boldsymbol{\Lambda}_{\mathbf{w}}$.

\section{Average Achievable rate maximization}

\label{average-rate}

In this section, we maximize the average achievable rate  under a constraint on the average RIS transmit power. To this end, a concave lower bound of the non-concave objective function is  constructed, and a KKT-based AO algorithm is developed.  

\subsection{Problem Formulation}

The proposed problem can be formulated as follows
\begin{subequations}
\label{Pro:max-}
\begin{align}
\mathop{\max}\limits _{\mathbf{F},\mathbf{w}} & \thinspace\thinspace\sum_{k=1}^{K}\mathbb{E}_{\mathbf{h}_{\mathrm{r},k}|{\bf G}_k}\left\{ R_{k}({\bf F},{\bf w})\right\} \label{eq:mi3}\\
\textrm{s.t.} & \thinspace\thinspace||\mathbf{F}||_{F}^{2}\leq P_{N},\label{eq:min-3}\\
 & \thinspace\thinspace\mathbb{E}_{\mathbf{H_{\mathrm{dr}}}|{\bf G}_k}\left\{ ||\boldsymbol{\Lambda}_{\mathbf{w}}\mathbf{H_{\mathrm{dr}}}{\bf F}||_{F}^{2}+||\mathbf{w}||_{F}^{2}\sigma_{z}^{2}\right\} \leq P_{M},\label{eq:mi2}\\
 & \thinspace\thinspace1\leq|w_{m}|^{2}\leq a_{max},\forall m,\label{eq:f3}
\end{align}
\end{subequations}
where $P_{M}$ is the maximum average RIS transmit power. 

Since $\mathbb{E}_{\mathbf{h}_{\mathrm{r},k}|{\bf G}_k}\left\{ R_{k}({\bf F},{\bf w})\right\}$ and $\mathbb{E}_{\mathbf{H_{\mathrm{dr}}}|{\bf G}_k}\left\{ ||\boldsymbol{\Lambda}_{\mathbf{w}}\mathbf{H_{\mathrm{dr}}}{\bf F}||_{F}^{2}+||\mathbf{w}||_{F}^{2}\sigma_{z}^{2}\right\} $ are not analyticaly tractable, Problem (\ref{Pro:max-}) cannot be solved directly. Thus, based on (\ref{eq:frfr}) and (\ref{eq:rate-a}), Problem (\ref{Pro:max-}) is lower bounded as follows
\begin{subequations}
\label{Pro:maxrate-1}
\begin{align}
\mathop{\max}\limits _{\mathbf{F},\mathbf{w}} & \thinspace\thinspace\sum_{k=1}^{K}\bar{R}_{k}\left(\mathbf{F},\mathbf{w}\right)\label{eq:min-power-obj-2-1}\\
\textrm{s.t.} & \thinspace\thinspace||\mathbf{F}||_{F}^{2}\leq P_{N},\label{eq:min-power-cons1-1-1-1-2}\\
 & \thinspace\thinspace P({\bf F},{\bf w})\leq P_{M},\label{eq:min-power-cons2-2-1-1}\\
 & \thinspace\thinspace1\leq|w_{m}|^{2}\leq a_{max},\forall m.\label{eq:re-1}
\end{align}
\end{subequations}
Problem (\ref{Pro:maxrate-1}) is still difficult to solve due to the non-concave
objective function in (\ref{eq:min-power-obj-2-1}), the non-convex amplification gain constraints in (\ref{eq:re-1}), and the coupling of variables
$\mathbf{F}$ and $\mathbf{w}$.

\subsection{Problem Reformulation}

In the following, we propose an AO algorithm to solve Problem (\ref{Pro:maxrate-1})
based on the MM algorithm (see, e.g., \cite{Hunter2004MM,MM} for 
tutorial introductions to  MM algorithms). Specifically, the key
idea of  MM algorithms is to construct an easy-to-solve surrogate
problem by deriving a minorizer of the original non-convex objective
function,  which is then used for optimization. Specifically,
assuming that $f(\mathbf{x})$ is the original objective function which
needs to be maximized over a convex set $\mathcal{S}_{x}$,  its
minorizers (denoted by $\widetilde{f}(\mathbf{x}|\mathbf{x}^{n})$)
at a given point $\mathbf{x}^{n}$ should satisfy the following conditions
\cite{MM}:
\begin{align*}
\mathrm{(A1):} & \widetilde{f}(\mathbf{x}^{n}|\mathbf{x}^{n})=f(\mathbf{x}^{n}),\forall\mathbf{x}^{n}\in\mathcal{S}_{x};\\
\mathrm{(A2):} & \widetilde{f}(\mathbf{x}|\mathbf{x}^{n})\leq f(\mathbf{x}),\forall\mathbf{x},\mathbf{x}^{n}\in\mathcal{S}_{x};\\
\mathrm{(A3):} & \widetilde{f}^{'}(\mathbf{x}|\mathbf{x}^{n};\mathbf{d})|_{\mathbf{x}=\mathbf{x}^{n}}=f^{'}(\mathbf{x}^{n};\mathbf{d}),\forall\mathbf{d}\thinspace\thinspace\mathrm{\textrm{with}}\thinspace\thinspace\mathbf{x}^{n}+\mathbf{d}\in\mathcal{S}_{x};\\
\mathrm{(A4):} & \widetilde{f}(\mathbf{x}|\mathbf{x}^{n})\thinspace\thinspace\textrm{is continuous in \ensuremath{\mathbf{x}} and \ensuremath{\mathbf{x}^{n}}, }
\end{align*}
where $f^{'}(\mathbf{x}^{n};\mathbf{d})$, defined as the direction
derivative of $f(\mathbf{x}^{n})$ in  direction $\mathbf{d}$,
is given by
\[
f^{'}(\mathbf{x}^{n};\mathbf{d})=\underset{\kappa\rightarrow0}{\textrm{lim}}\frac{f(\mathbf{x}^{n}+\kappa\mathbf{d})-f(\mathbf{x}^{n})}{\kappa}.
\]

Based on the MM framework, we derive a quadratic lower bound of $\bar{R}_{k}\left(\mathbf{F},\mathbf{w}\right)$
shown in the following lemma, the proof of which is similar to the proof in 
  \cite[Appendix A]{Gui2019IRS}.

\begin{lemma}\label{lemma-1} For a fixed point $\{\mathbf{F}^{n},\mathbf{w}^{n}\}$,
$\bar{R}_{k}\left(\mathbf{F},\mathbf{w}\right)$ is minorized by the
concave surrogate function $\widetilde{R}_{k}\left(\mathbf{F},\mathbf{w}|\mathbf{F}^{n},\mathbf{w}^{n}\right)$ given by
\begin{align}
 \widetilde{R}_{k}\left(\mathbf{F},\mathbf{w}|\mathbf{F}^{n},\mathbf{w}^{n}\right)
  =\textrm{const}_{k}+2\mathit{\mathrm{Re}}\left\{ a_{k}\tilde{\mathbf{w}}^{\mathrm{H}}\mathbf{H}_{k}{\bf f}_{k}\right\} -b_{k}(||\tilde{\mathbf{w}}^{\mathrm{H}}\mathbf{H}_{k}{\bf F}||_{2}^{2}+\mathbf{w}^{\mathrm{H}}\boldsymbol{\Psi}_{k}\mathbf{w}),\label{eq:t5}
\end{align}
where
\begin{align*}
 & a_{k}=\frac{t_{k}^{n,*}}{r_{k}^{n}-|t_{k}^{n}|^{2}},  \;\;\;\;\;\;\; b_{k}=\frac{|t_{k}^{n}|^{2}}{r_{k}^{n}(r_{k}^{n}-|t_{k}^{n}|^{2})},  \;\;\;\;\;\;\;
 \textrm{const}_{k}=R_{k}\left(\mathbf{F}^{n},\mathbf{w}^{n}\right)-b_{k}(\sigma_{k}^{2}+r_{k}^{n}),\\
 & t_{k}^{n}=(\tilde{\mathbf{w}}^{n})^{\mathrm{H}}\mathbf{H}_{k}{\bf f}_{k}^{n}, \;\;\;\;\;
 r_{k}^{n}=\sum_{i=1}^{K}|(\tilde{\mathbf{w}}^{n})^{\mathrm{H}}\mathbf{H}_{k}{\bf f}_{i}^{n}|^{2}+(\tilde{\mathbf{w}}^{n})^{\mathrm{H}}\boldsymbol{\Psi}_{k}\mathbf{w}^{n}+\sigma_{k}^{2}.
\end{align*}
\end{lemma}

Function (\ref{eq:t5}) is  biconcave in  $\mathbf{F}$ and $\mathbf{w}$, which motivates us to update $\mathbf{F}$ and $\mathbf{w}$ in an iterative manner. In particular, in the proposed AO algorithm, we first update  $\mathbf{F}$ based on the concave function $\widetilde{R}_{k}\left(\mathbf{F}|\mathbf{F}^{n}\right)=\widetilde{R}_{k}\left(\mathbf{F},\mathbf{w}|\mathbf{F}^{n},\mathbf{w}^{n}\right)$ for a given $\mathbf{w}$, and then we update $\mathbf{w}$ based on the concave function $\widetilde{R}_{k}\left(\mathbf{w}|\mathbf{w}^{n}\right)=\widetilde{R}_{k}\left(\mathbf{F},\mathbf{w}|\mathbf{F}^{n},\mathbf{w}^{n}\right)$ for a given $\mathbf{F}$.

\subsection{Optimization of Precoding Matrix ${\bf F}$}

By using Lemma \ref{lemma-1}, a lower bound  of the objective function in (\ref{eq:min-power-obj-2-1}) with respect to  ${\bf F}$, denoted
by $\widetilde{R}_{sum}(\mathbf{F})$, is obtained as
\begin{align}
\widetilde{R}_{sum}(\mathbf{F}) & =\sum_{k=1}^{K}\widetilde{R}_{k}\left(\mathbf{F},\mathbf{w}|\mathbf{F}^{n},\mathbf{w}^{n}\right)\nonumber \\
 & =\textrm{const}_{F}+2\mathit{\mathrm{Re}}\left\{ \mathrm{Tr}\left\{ \mathbf{C}_{F}^{\mathrm{H}}{\bf F}\right\} \right\} -\mathrm{Tr}\left\{ {\bf F}^{\mathrm{H}}\mathbf{A}_{F}{\bf F}\right\} ,\label{eq:lo}
\end{align}
where $\textrm{const}_{F} =\sum_{k=1}^{K}\textrm{const}_{k}-\mathbf{w}^{\mathrm{H}}(\sum_{k=1}^{K}b_{k}\boldsymbol{\Psi}_{k})\mathbf{w}$, $\mathbf{C}_{F}  =\sum_{k=1}^{K}a_{k}^{*}\mathbf{H}_{k}^{\mathrm{H}}\tilde{\mathbf{w}}\mathbf{t}_{k}^{\mathrm{H}}$, $\mathbf{A}_{F}  =\sum_{k=1}^{K}b_{k}\mathbf{H}_{k}^{\mathrm{H}}\tilde{\mathbf{w}}\tilde{\mathbf{w}}^{\mathrm{H}}\mathbf{H}_{k}$, 
 and $\mathbf{t}_{k}\in\mathbb{R}^{K\times1}$ is a selection vector
in which the $k$-th element is equal to one and all the other elements
are equal to zero.

After some manipulations, the average RIS transmit power in (\ref{eq:frfr})
can be rewritten as a quadratic function of ${\bf F}$ as follows 
\begin{align}
P({\bf F})= & \frac{\beta_{0}\delta_{0}}{\delta_{0}+1}\mathrm{Tr}\left\{ {\bf F}^{\mathrm{H}}\bar{\mathbf{H}}_{\mathrm{dr}}^{\mathrm{H}}\boldsymbol{\Lambda}_{\mathrm{RIS}}^{a}\bar{\mathbf{H}}_{\mathrm{dr}}{\bf F}\right\} +\frac{\beta_{0}}{\delta_{0}+1}\mathrm{Tr}\left\{ \boldsymbol{\Lambda}_{\mathrm{RIS}}^{a}\boldsymbol{\Sigma}_{\mathrm{R}}\right\} \mathrm{Tr}\left\{ {\bf F}^{\mathrm{H}}\boldsymbol{\Sigma}_{\mathrm{B}}{\bf F}\right\} +||\mathbf{w}||_{2}^{2}\sigma_{z}^{2}\nonumber \\
= & \frac{\beta_{0}\delta_{0}}{\delta_{0}+1}\mathrm{Tr}\left\{ {\bf F}^{\mathrm{H}}\bar{\mathbf{H}}_{\mathrm{dr}}^{\mathrm{H}}\boldsymbol{\Lambda}_{\mathrm{RIS}}^{a}\bar{\mathbf{H}}_{\mathrm{dr}}{\bf F}\right\} +\frac{\beta_{0}}{\delta_{0}+1}\mathrm{Tr}\left\{ \boldsymbol{\Lambda}_{\mathrm{RIS}}^{a}\right\} \mathrm{Tr}\left\{ {\bf F}^{\mathrm{H}}\boldsymbol{\Sigma}_{\mathrm{B}}{\bf F}\right\} +||\mathbf{w}||_{2}^{2}\sigma_{z}^{2}\nonumber \\
= & \mathrm{Tr}\left\{ {\bf F}^{\mathrm{H}}{\bf D}_{F}{\bf F}\right\} +||\mathbf{w}||_{2}^{2}\sigma_{z}^{2},\label{eq:ji}
\end{align}
where we have $\mathrm{Tr}\left\{ \boldsymbol{\Lambda}_{\mathrm{RIS}}^{a}\boldsymbol{\Sigma}_{\mathrm{R}}\right\} =||\mathbf{w}||_{2}^{2}$
and ${\bf D}_{F}=\frac{\beta_{0}\delta_{0}}{\delta_{0}+1}\bar{\mathbf{H}}_{\mathrm{dr}}^{\mathrm{H}}\boldsymbol{\Lambda}_{\mathrm{RIS}}^{a}\bar{\mathbf{H}}_{\mathrm{dr}}+\frac{\beta_{0}}{\delta_{0}+1}||\mathbf{w}||_{2}^{2}\boldsymbol{\Sigma}_{\mathrm{B}}$.

Combining (\ref{eq:lo}) with (\ref{eq:ji}), and ignoring irrelevant constant terms, the surrogate subproblem of (\ref{Pro:maxrate-1}) with respect to
${\bf F}$ for a given ${\bf w}$ is formulated as follows
\begin{subequations}
\label{Pro:max-f}
\begin{align}
\mathop{\max}\limits _{\mathbf{F}} & \thinspace\thinspace2\mathit{\mathrm{Re}}\left\{ \mathrm{Tr}\left\{ \mathbf{C}_{F}^{\mathrm{H}}{\bf F}\right\} \right\} -\mathrm{Tr}\left\{ {\bf F}^{\mathrm{H}}\mathbf{A}_{F}{\bf F}\right\} \label{eq:obj-f1}\\
\textrm{s.t.} & \thinspace\thinspace||\mathbf{F}||_{F}^{2}\leq P_{N},\label{eq:max-f-cons1}\\
 & \thinspace\thinspace\mathrm{Tr}\left\{ {\bf F}^{\mathrm{H}}{\bf D}_{F}{\bf F}\right\} +||\mathbf{w}||_{2}^{2}\sigma_{z}^{2}\leq P_{M}.\label{eq:max-f-cons2}
\end{align}
\end{subequations}
 Problem (\ref{Pro:max-f}) is a standard second-order cone programming (SOCP) problem and can be solved with CVX. However, the computational complexity of SOCP-based algorithms is high. In the following, we solve  Problem (\ref{Pro:max-f})  by exploiting  the standard dual decomposition method. 
In particular, the Lagrange function of Problem (\ref{Pro:max-f})
is given by
\begin{align*}
 & \mathcal{L}(\mathbf{F},\gamma_{F},\mu_{F})\\
= & \mathrm{Tr}\left\{ {\bf F}^{\mathrm{H}}\mathbf{A}_{F}{\bf F}\right\} -2\mathit{\mathrm{Re}}\left\{ \mathrm{Tr}\left\{ \mathbf{C}_{F}^{\mathrm{H}}{\bf F}\right\} \right\} +\gamma_{F}(||\mathbf{F}||_{F}^{2}-P_{N})+\mu_{F}(\mathrm{Tr}\left\{ {\bf F}^{\mathrm{H}}{\bf D}_{F}{\bf F}\right\} +||\mathbf{w}||_{2}^{2}\sigma_{z}^{2}-P_{M}),
\end{align*}
where  Lagrange multipliers $\gamma_{F}\geq0$ and $\mu_{F}\geq0$
are associated with constraints (\ref{eq:max-f-cons1}) and (\ref{eq:max-f-cons2}),
respectively.  The dual function $d(\gamma_{F},\mu_{F})$ is given by
\begin{align}\label{dede}
	d(\gamma_{F},\mu_{F})=\mathop{\max}\limits _{\mathbf{F}} \mathcal{L}(\mathbf{F},\gamma_{F},\mu_{F}).
\end{align}
Thus, the dual problem of Problem (\ref{Pro:max-f}) can be formulated as follows
\begin{align}\label{okoko}
	\mathop{\min}\limits _{\gamma_{F}\geq0,\mu_{F}\geq0} d(\gamma_{F},\mu_{F}).
\end{align}
Firstly, by exploiting the first-order KKT necessary and sufficient
condition of the problem in (\ref{dede}), i.e., $\frac{\partial\mathcal{L}(\mathbf{F},\gamma_{F},\mu_{F})}{\partial\mathbf{F}^{*}}  =\left(\mathbf{A}_{F}+\gamma_{F}^{\rm opt}{\bf I}+\mu_{F}^{\rm opt}{\bf D}_{F}\right){\bf F}_{\rm opt}-\mathbf{C}_{F}=\mathbf{0}$, 
we  obtain the optimal solution of ${\bf F}$ for fixed dual variables $(\gamma_{F}^{[v]},\mu_{F}^{[v]})$ in iteration $v$ as follows
\begin{align}
\mathbf{F}(\gamma_{F}^{[v]},\mu_{F}^{[v]}) & =\left(\mathbf{A}_{F}+\gamma_{F}^{[v]}{\bf I}+\mu_{F}^{[v]}{\bf D}_{F}\right)^{-1}\mathbf{C}_{F}.\label{eq:f-opt}
\end{align}
Then, the dual problem (\ref{okoko}) can be solved by the gradient projection algorithm, i.e., the dual variables are updated as follows
\begin{subequations}\label{kokop}
\begin{align}
	\gamma_{F}^{[v+1]} & =\left[\gamma_{F}^{[v]}+\varsigma^{[v]}\nabla_{\gamma_{F}}d(\gamma_{F}^{[v]},\mu_{F}^{[v]})\right]^{+}, \\
	\mu_{F}^{[v+1]} & =\left[\mu_{F}^{[v]}+\varsigma^{[v]}\nabla_{\mu_{F}}d(\gamma_{F}^{[v]},\mu_{F}^{[v]})\right]^{+}, 
\end{align}
\end{subequations}
where 
\begin{subequations}\label{kokop}
	\begin{align}
		\nabla_{\gamma_{F}}d(\gamma_{F}^{[v]},\mu_{F}^{[v]}) & =||\mathbf{F}(\gamma_{F}^{[v]},\mu_{F}^{[v]})||_{F}^{2}-P_{N} \\
		\nabla_{\mu_{F}}d(\gamma_{F}^{[v]},\mu_{F}^{[v]}) & =\mathrm{Tr}\left\{ ({\bf F}(\gamma_{F}^{[v]},\mu_{F}^{[v]}))^{\mathrm{H}}{\bf D}_{F}{\bf F}(\gamma_{F}^{[v]},\mu_{F}^{[v]})\right\} +||\mathbf{w}||_{2}^{2}\sigma_{z}^{2}-P_{M}.
	\end{align}
\end{subequations}
The initialization points can be set as $\gamma_{F}^{[0]}=0$ and $\mu_{F}^{[0]}=0$.  Then, (\ref{eq:f-opt}) and (\ref{kokop}) are updated in an alternating manner until $||(\gamma_{F}^{[v+1]},\mu_{F}^{[v+1]})-(\gamma_{F}^{[v]},\mu_{F}^{[v]})||\rightarrow0$. Note that although the algorithm proposed for solving Problem (\ref{Pro:max-f}) requires iterations, the computational complexity is comparatively low   due to closed-form expressions employed in  each iteration,  see Section \ref{iii-complexity}.

\subsection{Optimization of Reflection Vector ${\bf w}$ }

In order to facilitate the subsequent derivations, we convert the surrogate objective function\\
 $\sum_{k=1}^{K}\widetilde{R}_{k}\left(\mathbf{F},\mathbf{w}|\mathbf{F}^{n},\mathbf{w}^{n}\right)$
and average RIS transmit power in (\ref{eq:frfr}) into  quadratic
functions of ${\bf w}$ as follows
\begin{align}
\widetilde{R}_{sum}(\mathbf{w}) & =\sum_{k=1}^{K}\widetilde{R}_{k}\left(\mathbf{F},\mathbf{w}|\mathbf{F}^{n},\mathbf{w}^{n}\right)\nonumber \\
 & =\textrm{const}_{w}+2\mathit{\mathrm{Re}}\left\{ \mathbf{w}^{\mathrm{H}}\mathbf{c}_{w}\right\} -\mathbf{w}^{\mathrm{H}}\mathbf{A}_{w}\mathbf{w},\label{eq:vf}
\end{align}
with $\textrm{const}_{w} =\sum_{k=1}^{K}\textrm{const}_{k}+2\mathit{\mathrm{Re}}\{ \sum_{k=1}^{K}a_{k}\mathbf{h}_{k}^{\mathrm{H}}{\bf f}_{k}\} -\sum_{k=1}^{K}b_{k}\mathbf{h}_{k}^{\mathrm{H}}{\bf F}{\bf F}^{\mathrm{H}}\mathbf{h}_{k}$, $\mathbf{c}_{w}  =\sum_{k=1}^{K}a_{k}\mathbf{G}_{k}{\bf f}_{k}-\\
\sum_{k=1}^{K}b_{k}\mathbf{G}_{k}{\bf F}{\bf F}^{\mathrm{H}}\mathbf{h}_{k}$, and  $\mathbf{A}_{w}  =\sum_{k=1}^{K}b_{k}\mathbf{G}_{k}{\bf F}{\bf F}^{\mathrm{H}}\mathbf{G}_{k}^{\mathrm{H}}+(\sum_{k=1}^{K}b_{k}\boldsymbol{\Psi}_{k})$,
and
\begin{align}
P(\mathbf{w}) & =\mathbf{w}^{\mathrm{H}}{\bf D}_{w}\mathbf{w},\label{eq:bf}
\end{align}
with  ${\bf D}_{w}=\frac{\beta_{0}\delta_{0}}{\delta_{0}+1}\mathrm{Diag}\left(\mathrm{diag}\left(\bar{\mathbf{H}}{\bf F}{\bf F}^{\mathrm{H}}\bar{\mathbf{H}}^{\mathrm{H}}\right)\right)+\frac{\beta_{0}}{\delta_{0}+1}\mathrm{Tr}\left\{ {\bf F}^{\mathrm{H}}\boldsymbol{\Sigma}_{\mathrm{B}}{\bf F}\right\} {\bf I}+\sigma_{z}^{2}{\bf I}$.

Exploiting (\ref{eq:vf}) and (\ref{eq:bf}),  we formulate a  surrogate subproblem for (\ref{Pro:maxrate-1}) with respect to ${\bf w}$ for a given  ${\bf F}$  as follows
\begin{subequations}
\label{Pro:max-rate-w}
\begin{align}
\mathop{\max}\limits _{\mathbf{w}} & \thinspace\thinspace2\mathit{\mathrm{Re}}\left\{ \mathbf{w}^{\mathrm{H}}\mathbf{c}_{w}\right\} -\mathbf{w}^{\mathrm{H}}\mathbf{A}_{w}\mathbf{w}\label{eq:max-rate-w}\\
\textrm{s.t.} & \thinspace\thinspace\mathbf{w}^{\mathrm{H}}{\bf D}_{w}\mathbf{w}\leq P_{M},\label{eq:c1-w}\\
 & \thinspace\thinspace1\leq|w_{m}|^{2}\leq a_{max},\forall m.\label{eq:c2-w}
\end{align}
\end{subequations}
Problem (\ref{Pro:max-rate-w})
can be transformed into an SOCP by relaxing the non-convex constraint
$1\leq|w_{m}|^{2}$ in (\ref{eq:c2-w}) via a  linear approximate
constraint $1\leq2{\rm Re}\{(w_{m}^{n})^{*}w_{m}\}-|w_{m}^{n}|^{2}$
by using the first-order Taylor approximation at fixed point $w_{m}^{n}$.
The resulting approximate SOCP problem is given by
\begin{subequations}
\label{Pro:max-exp-rate-w-1}
\begin{align}
\mathop{\max}\limits _{\mathbf{w}} & \thinspace\thinspace2\mathit{\mathrm{Re}}\left\{ \mathbf{w}^{\mathrm{H}}\mathbf{c}_{w}\right\} -\mathbf{w}^{\mathrm{H}}\mathbf{A}_{w}\mathbf{w}\label{eq:min-power-obj-5-1-1}\\
\textrm{s.t.} & \thinspace\thinspace\mathbf{w}^{\mathrm{H}}{\bf D}_{w}\mathbf{w}\leq P_{M},\label{eq:min-power-cons1-5-1-1}\\
 & \thinspace\thinspace1\leq2{\rm Re}\{(w_{m}^{n})^{*}w_{m}\}-|w_{m}^{n}|^{2},\forall m,\label{eq:min-power-cons2-5-1-1}\\
 & \thinspace\thinspace|w_{m}|^{2}\leq a_{max},\forall m.
\end{align}
\end{subequations}

To find a low-complexity solution for (\ref{Pro:max-exp-rate-w-1}), we adopt the alternating direction
method of multipliers (ADMM) \cite{boydADMM}. In particular, we introduce  auxiliary variable $\mathbf{u}=[u_{1},\cdots,u_{M}]^{T}$
such that $\mathbf{u}=\mathbf{w}$ and $1\leq|u_{m}|^{2}\leq a_{max},\forall m$.
The augmented Lagrangian of the optimization problem is given by
\[
\mathcal{L}_{\xi}(\mathbf{w},\mathbf{u},\boldsymbol{\eta})=\mathbf{w}^{\mathrm{H}}\mathbf{A}_{w}\mathbf{w}-2\mathit{\mathrm{Re}}\left\{ \mathbf{w}^{\mathrm{H}}\mathbf{c}_{w}\right\} +\zeta||\mathbf{w}-\mathbf{u}+\boldsymbol{\eta}||_{2}^{2},
\]
where $\zeta>0$ is a penalty parameter.   The benefit of including the penalty term is to make the dual function differentiable. The ADMM method comprises the following steps  \footnote{Please note that $w_m^n$ in Problem (\ref{Pro:max-exp-rate-w-1}) is the updated value in each iteration of the MM algorithm, while ${\bf w}^{[i+1]}$ in Problem (\ref{eq:nh}) is the updated value in each iteration of the ADMM method.}:
\begin{align}
\mathbf{w}^{[i+1]} & =\arg\min_{\mathbf{w}\in\{\mathbf{w}^{\mathrm{H}}{\bf D}_{w}\mathbf{w}\leq P_{M}\}}\,\,\mathcal{L}_{\xi}(\mathbf{w},\mathbf{u}^{[i]},\boldsymbol{\eta}^{[i]}),\label{eq:nh}\\
\mathbf{u}^{[i+1]} & =\arg\min_{\mathbf{u}\in\{1\leq|u_{m}|^{2}\leq a_{max},\forall m\}}\,\,\mathcal{L}_{\xi}(\mathbf{w}^{[i+1]},\mathbf{u},\boldsymbol{\eta}^{[i]}),\label{eq:nj}\\
\boldsymbol{\eta}^{[i+1]} & =\boldsymbol{\eta}^{[i]}+\mathbf{w}^{[i+1]}-\mathbf{u}^{[i+1]}.\label{eq:nk}
\end{align}

$\bullet$ Updating $\mathbf{w}$: $\mathbf{w}^{[i+1]}$ can be obtained using the
KKT conditions. We form the Lagrangian function of Problem (\ref{eq:nh})
with Lagrange multiplier $\gamma_{w}$ as
$
\mathcal{L}(\mathbf{w},\gamma_{w})=  \mathbf{w}^{\mathrm{H}}\mathbf{A}_{w}\mathbf{w}-2\mathit{\mathrm{Re}}\left\{ \mathbf{w}^{\mathrm{H}}\mathbf{c}_{w}\right\} +\zeta||\mathbf{w}-\mathbf{u}^{[i]}+\boldsymbol{\eta}^{[i]}||_{2}^{2}+\gamma_{w}(\mathbf{w}^{\mathrm{H}}{\bf D}_{w}\mathbf{w}-P_{M}),
$
and obtain the first-order KKT necessary condition for the optimal $\mathbf{w}^{[i+1]}$ as $\frac{\partial\mathcal{L}(\mathbf{w}^{[i+1]},\gamma_{w}^{\rm opt})}{\partial\mathbf{w}^{*}}= 
\left(\mathbf{A}_{w}+\gamma_{w}^{\rm opt}{\bf D}_{w}+\zeta\mathbf{I}\right)\mathbf{w}^{[i+1]}-\mathbf{c}_{w}+\zeta(\boldsymbol{\eta}^{[i]}-\mathbf{u}^{[i]})
=  \mathbf{0}$. 
Then, $\mathbf{w}^{[i+1]}$ is given by
\begin{align}
\mathbf{w}^{[i+1]}(\gamma_{w}^{\rm opt})=\left(\mathbf{A}_{w}+\gamma_{w}^{\rm opt}{\bf D}_{w}+\zeta\mathbf{I}\right)^{-1}(\mathbf{c}_{w}-\zeta(\boldsymbol{\eta}^{[i]}-\mathbf{u}^{[i]})).\label{opt-w-admm}
\end{align}
 Function $g_{w}(\gamma_{w})=(\mathbf{w}^{[i+1]}(\gamma_{w}))^{\mathrm{H}}{\bf D}_{w}\mathbf{w}^{[i+1]}(\gamma_{w})$
is a monotonically decreasing function of $\gamma_{w}$.
 If
 $g_{w}(0)\leq P_{M}$,
then $\mathbf{w}^{[i+1]}=\left(\mathbf{A}_{w}+\zeta\mathbf{I}\right)^{-1}(\mathbf{c}_{w}-\zeta(\boldsymbol{\eta}^{[i]}-\mathbf{u}^{[i]}))$.
Otherwise,  $g_{w}(0)> P_{M}$.  Based on the complementary condition of $\gamma_{w}(g_{w}(\gamma_{w})-P_{M})=0$, we need to find a positive $\gamma_{w}$ such that  $g_{w}(\gamma_{w})-P_{M}=0$. 
Defining $\gamma_{w,1}=\sqrt{\frac{(\mathbf{c}_{w}-\zeta(\boldsymbol{\eta}^{[i]}-\mathbf{u}^{[i]}))^{\mathrm{H}}{\bf D}_{w}^{-1}(\mathbf{c}_{w}-\zeta(\boldsymbol{\eta}^{[i]}-\mathbf{u}^{[i]}))}{P_{M}}}$, we have $g_{w}(\gamma_{w,1}) < \left(\gamma_{w,1}\right)^{-2} (\mathbf{c}_{w}-\zeta(\boldsymbol{\eta}^{[i]}-\mathbf{u}^{[i]}))^{\mathrm{H}}{\bf D}_{w}^{-1}(\mathbf{c}_{w}-\zeta(\boldsymbol{\eta}^{[i]}-\mathbf{u}^{[i]})) = P_{M}$. Then, a unique $\gamma_{w}^{\rm opt}\in(0,\gamma_{w,1})$ must exist such that $g_{w}(\gamma_{w}^{\rm opt})=P_{M}$, and thus $\gamma_{w}^{\rm opt}$ can be found by using a one-dimensional search.

$\bullet$ Updating $\mathbf{u}$: The optimization problem in (\ref{eq:nj})
is equivalent to
\begin{align}
\min\limits _{\mathbf{u}}  \thinspace\thinspace||\mathbf{w}-\mathbf{u}+\boldsymbol{\eta}||_{2}^{2}  \;\;\;\;
\textrm{s.t.}  \thinspace\thinspace1\leq|u_{m}|^{2}\leq a_{max},\forall m.\label{eq:c2-u}
\end{align}
 Its solution is given by $\mathbf{u}^{[i+1]}=[|\mathbf{w}^{[i+1]}+\boldsymbol{\eta}^{[i]}|]_{1}^{\sqrt{a_{max}}}\exp(j\angle(\mathbf{w}^{[i+1]}+\boldsymbol{\eta}^{[i]}))$,
where operator $|\cdot|$ returns the  elementwise absolute value and operator
$[\mathbf{x}]_{\underline{u}}^{\overline{u}}$ maps $\mathbf{x}$
elementwise onto the interval $[\underline{u},\overline{u}]$.

\subsection{Algorithm Development}

Under the MM framework, the solution $\mathbf{F}$ of Problem (\ref{Pro:max-f}) and the solution $\mathbf{w}$ of Problem (\ref{Pro:max-exp-rate-w-1}) in each AO iteration can be obtained with low complexity using the proposed KKT-based and ADMM methods, respectively.  The convergence speed of the MM algorithm will be affected by
the tightness of the lower bound  of the original objective function given
in Lemma \ref{lemma-1}. Thus, an acceleration method, called SQUAREM \cite{varadhan2008SQUAREM}, is adopted to accelerate the MM-based algorithm, as is summarized in Algorithm \ref{Algorithm-ao}. $F_{F}\left(\mathbf{F}^{n}\right)$ in
Step 9 and $F_{w}\left(\mathbf{w}^{n}\right)$ in Step 20 represent the
objective function values of Problem (\ref{Pro:max-f}) and Problem
(\ref{Pro:max-exp-rate-w-1}) in the $n$-th iteration, respectively.

$\mathcal{P}_{F}(\cdot)$ in Step 8 and $\mathcal{P}_{w}(\cdot)$  in Step 19 are projection operations onto the nonlinear constraint sets of $\mathbf{F}$ and $\mathbf{w}$,
respectively, which ensure the feasibility of the updated solutions. The projection operation is defined as  $\mathcal{P}(\mathbf{x})=\arg\min_{\mathbf{z}\in\mathcal{S}}||\mathbf{z}-\mathbf{x}||_{2}^{2}$, where $\mathcal{S}$ is the  constraint set of $\mathbf{z}$  \cite[Equ. (4.4.13)]{Xinda2017}. Therefore, for the power constraint set of $\mathbf{F}$, $\mathcal{P}_{F}(\cdot)$
is obtained as follows
\begin{align}
\mathcal{P}_{F}({\bf X})={\rm arg}\mathop{\min}\limits _{\mathbf{F}}  \thinspace\thinspace||{\bf F}-{\bf X}||_{F}   \;\;\;\;
\textrm{s.t.}  \thinspace\thinspace (\ref{eq:max-f-cons1}), (\ref{eq:max-f-cons2}).
\end{align}
and for the  power and amplification gain constraints of ${\bf w}$,
$\mathcal{P}_{w}(\cdot)$ is obtained as follows
\begin{align}
\mathcal{P}_{w}({\bf x})={\rm arg}\mathop{\min}\limits _{\mathbf{w}}  \thinspace\thinspace||{\bf w}-{\bf x}||_{2}      \;\;\;\;
\textrm{s.t.}  \thinspace\thinspace (\ref{eq:c1-w}), (\ref{eq:c2-w}).
\end{align}
 Steps 9 to 12 and Steps 20 to 23 are used to maintain the monotonicity of the objective function values.

\begin{algorithm}
\caption{Low-complexity MM algorithm}
\label{Algorithm-ao} \begin{algorithmic}[1] \REQUIRE Initialize
$\mathbf{F}^{0}$ and $\mathbf{w}^{0}$. Set $n=1$

\REPEAT

\STATE Set $\mathbf{w}=\mathbf{w}^{n-1}$
\STATE Obtain $\mathbf{F}_{1}$ from Problem (\ref{Pro:max-f}) based on $\mathbf{F}^{n-1}$ 
\STATE
Obtain $\mathbf{F}_{2}$ from Problem (\ref{Pro:max-f}) based on $\mathbf{F}_{1}$ 
\STATE $\mathbf{R}_{1}=\mathbf{F}_{1}-\mathbf{F}^{n-1}$
\STATE $\mathbf{R}_{2}=\mathbf{F}_{2}-\mathbf{F}_{1}-\mathbf{R}_{1}$
\STATE $\omega_{F}=-\frac{||\mathbf{R}_{1}||_{F}}{||\mathbf{R}_{2}||_{F}}$
\STATE $\mathbf{F}^{n}=-\mathcal{P}_{F}(\mathbf{F}^{n-1}-2\omega_{F}\mathbf{R}_{1}+\omega_{F}^{2}\mathbf{R}_{2})$
\WHILE {$F_{F}(\mathbf{F}^{n})<F_{F}(\mathbf{F}^{n-1})$}
\STATE $\;\;\;\;\omega_{F}=(\omega_{F}-1)/2$
\STATE $\;\;\;\;\mathbf{F}^{n}=-\mathcal{P}_{F}(\mathbf{F}^{n-1}-2\omega_{F}\mathbf{R}_{1}+\omega_{F}^{2}\mathbf{R}_{2})$

\ENDWHILE

\STATE Set $\mathbf{F}=\mathbf{F}^{n}$
 \STATE Obtain $\mathbf{w}_{1}$
from Problem (\ref{Pro:max-exp-rate-w-1}) based on $\mathbf{w}^{n-1}$
\STATE Obtain $\mathbf{w}_{2}$ from Problem (\ref{Pro:max-exp-rate-w-1})
based on $\mathbf{w}^{1}$
\STATE $\mathbf{r}_{1}=\mathbf{w}_{1}-\mathbf{w}^{n-1}$
\STATE $\mathbf{r}_{2}=\mathbf{w}_{2}-\mathbf{w}_{1}-\mathbf{r}_{1}$
\STATE $\omega_{w}=-\frac{||\mathbf{r}_{1}||_{2}}{||\mathbf{r}_{2}||_{2}}$
\STATE $\mathbf{w}^{n}=-\mathcal{P}_{w}(\mathbf{w}^{n-1}-2\omega_{w}\mathbf{r}_{1}+\omega_{w}^{2}\mathbf{r}_{2})$

\WHILE {$F_{w}\left(\mathbf{w}^{n}\right)<F_{w}\left(\mathbf{w}^{n-1}\right)|$}
\STATE $\;\;\;\;\omega_{w}=(\omega_{w}-1)/2$
\STATE $\;\;\;\;\mathbf{w}^{n}=-\mathcal{P}_{w}(\mathbf{w}^{n-1}-2\omega_{w}\mathbf{r}_{1}+\omega_{w}^{2}\mathbf{r}_{2})$

\ENDWHILE

\STATE $n=n+1$

\UNTIL $|F_{F}\left(\mathbf{F}^{n}\right)-F_{F}\left(\mathbf{F}^{n-1}\right)|\rightarrow0$
and $|F_{w}\left(\mathbf{w}^{n}\right)-F_{w}\left(\mathbf{w}^{n-1}\right)|\rightarrow0$

\end{algorithmic}
\end{algorithm}

\subsection{Complexity Analysis}
\label{iii-complexity}
Algorithm \ref{Algorithm-ao} requires  solving Problem (\ref{Pro:max-f}) and Problem (\ref{Pro:max-exp-rate-w-1}).  In the following complexity analysis, we neglect terms with low-order complexity. To solve Problem (\ref{Pro:max-f}),
we first need to calculate ${\bf A}_{F}$ and ${\bf D}_{F}$, which have computational complexity orders of $\mathcal{O}(M^{2}NK+MNK)$
and $\mathcal{O}(M^{2}NK)$, respectively. The calculation of ${\bf C}_{F}$ is similar to that of ${\bf A}_{F}$. Then, the inverse operation in (\ref{eq:f-opt})
has  complexity  $\mathcal{O}(M^{3})$. Therefore, the approximate computational
complexity of solving Problem (\ref{Pro:max-f}) is $\mathcal{O}(M^{3}+M^{2}NK+MNK)$,
where  constant coefficients are ignored. The computational complexity of
the ADMM algorithm used for solving Problem (\ref{Pro:max-exp-rate-w-1}) is mainly
determined by the calculation of ${\bf A}_{w}$, ${\bf D}_{w}$, and the inverse operation in (\ref{opt-w-admm}), which have complexities of  $\mathcal{O}(MNK^{2}+M^{2}NK)$, $\mathcal{O}(MNK)$, and $\mathcal{O}(M^{3})$, respectively. Also,  computing ${\bf c}_{w}$ involves similar steps as computing ${\bf A}_{w}$. Neglecting the constant coefficients, the approximate complexity of the ADMM algorithm is given by  $\mathcal{O}(M^{3}+MNK^{2}+M^{2}NK+MNK)$. Thus, the approximate complexity of Algorithm \ref{Algorithm-ao} per iteration is $\mathcal{O}(M^{3}+MNK^{2}+M^{2}NK+MNK)$.

\subsection{Convergence Analysis}
\label{Convergence Analysis}

Next, we analyze the convergence of the proposed algorithm. The monotonic convergence
of the MM algorithm has been proved in \cite{MM} and \cite{jacobson2007MM}.
In the following, we prove the monotonic convergence of Algorithm \ref{Algorithm-ao}.
Let $f(\mathbf{F},\mathbf{w})=\sum_{k=1}^{K}\bar{R}_{k}\left(\mathbf{F},\mathbf{w}\right)$
denote the objective value of Problem (\ref{Pro:maxrate-1}) and $\widetilde{f}(\mathbf{F},\mathbf{w})=\sum_{k=1}^{K}\widetilde{R}_{k}\left(\mathbf{F},\mathbf{w}\right)$
represent its minorizer. In the $n^{\mathrm{th}}$ iteration, 
given $\mathbf{w}^{n}$, we have
\[
f(\mathbf{F}^{n},\mathbf{w}^{n})=\widetilde{f}(\mathbf{F}^{n},\mathbf{F}^{n})\leq\widetilde{f}(\mathbf{F}^{n+1},\mathbf{F}^{n})\leq f(\mathbf{F}^{n+1},\mathbf{w}^{n}),
\]
where the first equality follows from condition (A1), the first inequality
is due to the optimal solution of Problem (\ref{Pro:max-f}), and
the second inequality follows from condition (A2). Subsequently, 
given $\mathbf{F}^{n+1}$, it is straightforward to show that
\[
f(\mathbf{F}^{n+1},\mathbf{w}^{n})=\widetilde{f}(\mathbf{w}^{n},\mathbf{w}^{n})\leq\widetilde{f}(\mathbf{w}^{n+1},\mathbf{w}^{n})\leq f(\mathbf{F}^{n+1},\mathbf{w}^{n+1}).
\]
Therefore, the sequence of objective values $\{f(\mathbf{F}^{n+1},\mathbf{w}^{n+1})\}$ generated by the  AO algorithm is monotonically non-decreasing.  Since $\mathbf{F}$ belongs to a convex set, every limit point of $\mathbf{F}^{n}$ is a d-stationay point of Problem (\ref{Pro:maxrate-1}) \cite{Gui2019IRS}. Furthermore, since $\mathbf{w}$ belongs to a non-convex set, every limit point of $\mathbf{w}^{n}$ is a B-stationay point of Problem (\ref{Pro:maxrate-1}) \cite{Gui2019IRS}.

\section{Outage constrained average power minimization}

\label{average-power}

In the previous section, we have  investigated the average rate maximization problem
for the  practical case where only partial CSI of the RIS-aided channels is available. 
 However,  this problem formulation cannot guarantee the QoS of the individual users and outages may occur in an uncontrolled manner. Thus, in order to ensure the QoS of the individual users,  in this section, we jointly optimize the beamforming matrices at both the BS and the RIS to guarantee  that the probability that the instantaneous achievable rate of each user  exceeds a target rate is larger than a predefined value, while minimizing the total transmit power consumption comprising the BS transmit power and the average RIS transmit power.  
  To obtain a tractable problem formulation,  the outage probability constraint is approximated by the BTI, and then an SDR-based AO algorithm is proposed to optimize the beamforming matrices.

 \subsection{Problem Formulation}
 
 The proposed  optimization problem is formulated as follows 
\begin{subequations}
\label{Pro:min-p2}
\begin{align}
\mathop{\min}\limits _{\mathbf{F},\mathbf{w}} & \thinspace\thinspace||\mathbf{F}||_{F}^{2}+\mathbb{E}_{\mathbf{H_{\mathrm{dr}}}|{\bf G}_k}\left\{ ||\boldsymbol{\Lambda}_{\mathbf{w}}\mathbf{H_{\mathrm{dr}}}{\bf F}||_{F}^{2}\right\} +||\boldsymbol{\Lambda}_{\mathbf{w}}||_{F}^{2}\sigma_{z}^{2}\label{eq:min-p4}\\
\textrm{s.t.} & \thinspace\thinspace\mathrm{Pr}\{R_{k}({\bf F},{\bf w})\geq r_{k}\}\geq1-\rho_{k},\forall k,\label{eq:min-p5}\\
 & \thinspace\thinspace1\leq|w_{m}|^{2}\leq a_{max},\forall m,
\end{align}
\end{subequations}
 where (\ref{eq:min-p5}) ensures that the probability that each user can successfully decode its message for a data rate of $r_{k}$ is no less than $1-\rho_{k}$, where $\rho_{1},\cdots,\rho_{K}\in(0,1]$
are the corresponding maximum outage probabilities.

\subsection{Problem Reformulation}

Problem (\ref{Pro:min-p2}) is computationally intractable since  outage probability constraint (\ref{eq:min-p5}) does not have a analytical expression. Therefore, we safely approximate   (\ref{eq:min-p5}) by some easy-to-handle constraints  by exploiting the following lemma.

\begin{lemma}\label{Bernstein-Type Inequalities} (Bernstein-Type
Inequality: Lemma 1 in \cite{W-K-MA2014}) Assume $f(\mathbf{x})=\mathbf{x}^{\mathrm{H}}\mathbf{U}\mathbf{x}+2\mathrm{Re}\{\mathbf{u}^{\mathrm{H}}\mathbf{x}\}+u$,
where $\mathbf{U}\in\mathbb{H}^{n\times n}$, $\mathbf{u}\in\mathbb{C}^{n\times1}$,
$u\in\mathbb{R}$, and $\mathbf{x}\in\mathbb{C}^{n\times1}\sim\mathcal{CN}(\mathbf{0},\mathbf{I})$.
Then, for any $\rho\in[0,1]$, the following approximation holds:
\begin{subequations}
\label{eq:Outage-5}
\begin{align}
 & \mathrm{Pr}\{\mathbf{x}^{\mathrm{H}}\mathbf{U}\mathbf{x}+2\mathrm{Re}\{\mathbf{u}^{\mathrm{H}}\mathbf{x}\}+u\leq0\}\geq1-\rho\label{eq:ml}\\
\Rightarrow & \mathrm{Tr}\left\{ \mathbf{U}\right\} +\sqrt{2\ln(1/\rho)}x-\ln(\rho)\lambda_{\max}^{+}(\mathbf{U})+u\leq0\\
\Rightarrow & \left\{ \begin{array}{c}
\mathrm{Tr}\left\{ \mathbf{U}\right\} +\sqrt{2\ln(1/\rho)}x-\ln(\rho)y+u\leq0\\
\sqrt{||\mathbf{U}||_{F}^{2}+2||\mathbf{u}||_{2}^{2}}\leq x\\
y\mathbf{I}-\mathbf{U}\succeq\mathbf{0},y\geq0,
\end{array}\right.
\end{align}
where $\lambda_{\max}^{+}(\mathbf{U})=\max(\lambda_{\max}(\mathbf{U}),0)$ and $\lambda_{\max}(\mathbf{U})$ denotes the maximum eigenvalue of $\bf U$.
$x$ and $y$ are slack variables.
\end{subequations}
 \end{lemma} Please refer to \cite{W-K-MA2014} for a proof of Lemma \ref{Bernstein-Type Inequalities}.

To utilize Lemma \ref{Bernstein-Type Inequalities},  the outage probability  in (\ref{eq:min-p5}) is firstly reformulated into the form of (\ref{eq:ml}):
\begin{align}
\mathrm{Pr}\left\{ R_{k}({\bf F},{\bf w})\geq r_{k}\right\} =\mathrm{Pr}\left\{ \tilde{\mathbf{w}}^{\mathrm{H}}\mathbf{H}_{k}\boldsymbol{\Phi}_{k}\mathbf{H}_{k}^{\mathrm{H}}\tilde{\mathbf{w}}-||\mathbf{h}_{\mathrm{r},k}^{\mathrm{H}}\boldsymbol{\Lambda}_{\mathbf{w}}||_{2}^{2}\sigma_{z}^{2}-\sigma_{k}^{2}\geq0\right\} ,\label{eq:o0}
\end{align}
where
\begin{subequations}
\begin{align}
\boldsymbol{\Phi}_{k}&={\bf f}_{k}\mathbf{f}_{k}^{\mathrm{H}}/(2^{r_{k}}-1)-{\bf F}_{-k}\mathbf{F}_{-k}^{\mathrm{H}},\label{eq:o9}\\
{\bf F}_{-k}&=[{\bf f}_{1},...,{\bf f}_{k},{\bf f}_{k+1},...,{\bf f}_{K}].
\end{align}
\end{subequations}
Substituting (\ref{eq:tr}) and $\tilde{\mathbf{h}}_{\mathrm{r},k}=\boldsymbol{\Sigma}_{\mathrm{r},k}^{1/2}\mathbf{e}_{\mathrm{r},k}$
into (\ref{eq:o0}), we have
\begin{align*}
 & \mathrm{Pr}\left\{ R_{k}({\bf F},{\bf w})\geq r_{k}\right\} =\mathrm{Pr}\left\{ \mathbf{e}_{\mathrm{r},k}^{\mathrm{H}}\mathbf{U}_{k}\mathbf{e}_{\mathrm{r},k}+2\mathrm{Re}\{\mathbf{u}_{k}^{\mathrm{H}}\mathbf{e}_{\mathrm{r},k}\}+u_{k}\leq0\right\} ,
\end{align*}
where
\begin{subequations}
\begin{align}
\mathbf{U}_{k} & =\frac{\beta_{k}}{\delta_{k}+1}\boldsymbol{\Sigma}_{\mathrm{r},k}^{1/2}\boldsymbol{\Lambda}_{\mathrm{RIS}}^{a}\boldsymbol{\Sigma}_{\mathrm{r},k}^{1/2},\\
\mathbf{u}_{k} & =\frac{\beta_{k}\sqrt{\delta_{k}}}{\delta_{k}+1}\boldsymbol{\Sigma}_{\mathrm{r},k}^{1/2}\boldsymbol{\Lambda}_{\mathrm{RIS}}^{a}\bar{\mathbf{h}}_{\mathrm{r},k},\\
u_{k} & =\frac{\beta_{k}\delta_{k}}{\delta_{k}+1}\bar{\mathbf{h}}_{\mathrm{r},k}^{\mathrm{H}}\boldsymbol{\Lambda}_{\mathrm{RIS}}^{a}\bar{\mathbf{h}}_{\mathrm{r},k}-\frac{1}{\sigma_{z}^{2}}\left(\tilde{\mathbf{w}}^{\mathrm{H}}\mathbf{H}_{k}\boldsymbol{\Phi}_{k}\mathbf{H}_{k}^{\mathrm{H}}\tilde{\mathbf{w}}-\sigma_{k}^{2}\right).\label{eq:c}
\end{align}
\end{subequations}

Furthermore,   the following theorem is provided to facilitate the subsequent derivations.
\begin{theorem}\label{pro1} Given matrices $\mathbf{A}\in\mathbb{C}^{N\times N}$,
	$\mathbf{b}\in\mathbb{C}^{N}$, and $\mathbf{C}\in\mathbb{C}^{N\times N}$,
	we have
	\begin{equation}
		\mathrm{Tr}\left\{ \mathbf{A}\mathrm{Diag}(\mathbf{b})\mathbf{C}\mathrm{Diag}(\mathbf{b})\right\} =\mathbf{b}^{\mathrm{T}}(\mathbf{A}^{\mathrm{T}}\odot\mathbf{C})\mathbf{b}.\label{eq:l0}
	\end{equation}
	
\end{theorem}\textbf{\textit{Proof: }}Please refer to Appendix \ref{appenxi1}.\hspace{11cm}$\blacksquare$

Then,  we establish the following identities: 
\begin{subequations}
\label{eq:Outage-7}
\begin{align}
\mathrm{Tr}\left\{ \mathbf{U}_{k}\right\}  & =\frac{\beta_{k}}{\delta_{k}+1}\mathrm{Tr}\left\{ \boldsymbol{\Lambda}_{\mathrm{RIS}}^{a}\right\} ,\label{eq:x2}\\
||\mathbf{U}_{k}||_{F}^{2} & =\left(\frac{\beta_{k}}{\delta_{k}+1}\right)^{2}\mathrm{Tr}\left\{ \boldsymbol{\Sigma}_{\mathrm{r},k}\boldsymbol{\Lambda}_{\mathrm{RIS}}^{a}\boldsymbol{\Sigma}_{\mathrm{r},k}\boldsymbol{\Lambda}_{\mathrm{RIS}}^{a}\right\} \nonumber \\
 & =\left(\frac{\beta_{k}}{\delta_{k}+1}\right)^{2}\mathbf{p}_{\mathrm{RIS}}^{\mathrm{T}}(\boldsymbol{\Sigma}_{\mathrm{r},k}^{\mathrm{T}}\odot\boldsymbol{\Sigma}_{\mathrm{r},k})\mathbf{p}_{\mathrm{RIS}},\label{eq:x3}\\
||\mathbf{u}_{k}||_{2}^{2} & =\left(\frac{\beta_{k}\sqrt{\delta_{k}}}{\delta_{k}+1}\right)^{2}\mathrm{Tr}\left\{ \bar{\mathbf{h}}_{\mathrm{r},k}\bar{\mathbf{h}}_{\mathrm{r},k}^{\mathrm{H}}\boldsymbol{\Lambda}_{\mathrm{RIS}}^{a}\boldsymbol{\Sigma}_{\mathrm{r},k}\boldsymbol{\Lambda}_{\mathrm{RIS}}^{a}\right\} \nonumber \\
 & =\left(\frac{\beta_{k}\sqrt{\delta_{k}}}{\delta_{k}+1}\right)^{2}\mathbf{p}_{\mathrm{RIS}}^{\mathrm{T}}((\bar{\mathbf{h}}_{\mathrm{r},k}^{*}\bar{\mathbf{h}}_{\mathrm{r},k}^{\mathrm{T}})\odot\boldsymbol{\Sigma}_{\mathrm{r},k})\mathbf{p}_{\mathrm{RIS}},\label{eq:x4}\\
\lambda(\mathbf{U}_{k}) & =\lambda\left(\frac{\beta_{k}}{\delta_{k}+1}\boldsymbol{\Sigma}_{\mathrm{r},k}^{1/2}\boldsymbol{\Lambda}_{\mathrm{RIS}}^{a}\boldsymbol{\Sigma}_{\mathrm{r},k}^{1/2}\right)\nonumber \\
 & =\frac{\beta_{k}}{\delta_{k}+1}\lambda(\boldsymbol{\Sigma}_{\mathrm{r},k}\boldsymbol{\Lambda}_{\mathrm{RIS}}^{a}),
\end{align}
\end{subequations}
 where $\mathbf{p}_{\mathrm{RIS}}=\mathrm{diag}(\boldsymbol{\Lambda}_{\mathrm{RIS}}^{a})$.
Equations (\ref{eq:x3}) and (\ref{eq:x4}) are obtained based on Theorem \ref{pro1}.

Therefore, applying Lemma \ref{Bernstein-Type Inequalities}, constraint (\ref{eq:min-p5}) can be approximated as follows
\begin{align}
 & \left\{ \begin{array}{c}
\frac{\beta_{k}}{\delta_{k}+1}\mathrm{Tr}\left\{ \boldsymbol{\Lambda}_{\mathrm{RIS}}^{a}\right\} +\sqrt{2\ln(1/\rho_{k})}x_{k}-\ln(\rho_{k})y_{k}+\frac{\beta_{k}\delta_{k}}{\delta_{k}+1}\bar{\mathbf{h}}_{\mathrm{r},k}^{\mathrm{H}}\boldsymbol{\Lambda}_{\mathrm{RIS}}^{a}\bar{\mathbf{h}}_{\mathrm{r},k}\\
-\frac{1}{\sigma_{z}^{2}}\left(\tilde{\mathbf{w}}^{\mathrm{H}}\mathbf{H}_{k}\boldsymbol{\Phi}_{k}\mathbf{H}_{k}^{\mathrm{H}}\tilde{\mathbf{w}}-\sigma_{k}^{2}\right)\leq0,\forall k,\\
\frac{\beta_{k}}{\delta_{k}+1}||\mathbf{C}_{k}^{1/2}\mathbf{p}_{\mathrm{RIS}}||\leq x_{k},\forall k,\\
y_{k}\mathbf{I}-\frac{\beta_{k}}{\delta_{k}+1}\boldsymbol{\Sigma}_{\mathrm{r},k}\boldsymbol{\Lambda}_{\mathrm{RIS}}^{a}\succeq\mathbf{0},\forall k,\\
y_{k}\geq0,\forall k,
\end{array}\right.\label{eq:BTI-1}
\end{align}
where $\mathbf{C}_{k}=\left(\boldsymbol{\Sigma}_{\mathrm{r},k}^{\mathrm{T}}+2\delta_{k}(\bar{\mathbf{h}}_{\mathrm{r},k}^{*}\bar{\mathbf{h}}_{\mathrm{r},k}^{\mathrm{T}})\right)\odot\boldsymbol{\Sigma}_{\mathrm{r},k}$, and $\mathbf{x}=[x_{1},\cdots,x_{K}]^{\mathrm{T}}$ and $\mathbf{y}=[y_{1},\cdots,y_{K}]^{\mathrm{T}}$ are  auxiliary variables.

Using $P({\bf F},{\bf w})$, defined in (\ref{eq:frfr}),  and the analytical constraints in (\ref{eq:BTI-1}), Problem (\ref{Pro:min-p2}) can be equivalently transformed into
\begin{subequations}
\label{Pro:min-power-1}
\begin{align}
\mathop{\min}\limits _{\mathbf{F},\mathbf{w},\mathbf{x},\mathbf{y}} & \thinspace\thinspace||\mathbf{F}||_{F}^{2}+P({\bf F},{\bf w})\label{eq:bi9}\\
\textrm{s.t.} & \thinspace\thinspace(\ref{eq:BTI-1}),\nonumber \\
 & \thinspace\thinspace1\leq|w_{m}|^{2}\leq a_{max},\forall m.
\end{align}
\end{subequations}

To overcome the coupling of variables $\mathbf{F}$ and $\mathbf{w}$,
we employ AO to solve Problem (\ref{Pro:min-power-1}).
The resulting non-convex subproblems for  $\mathbf{F}$ and $\mathbf{w}$
are separately relaxed by using  SDR  \cite{luo2010SDR}
and then solved with CVX in an iterative manner.

\subsection{Optimization of Precoding Matrix ${\bf F}$}

Given $\mathbf{w}$, we define new variables  $\boldsymbol{\Gamma}_{k}=\mathbf{f}_{k}\mathbf{f}_{k}^{\mathrm{H}}$ constrained by $\boldsymbol{\Gamma}_{k}\succeq\mathbf{0}$
	and $\mathrm{rank}(\boldsymbol{\Gamma}_{k})=1$, $\forall k$. Correspondingly,
 $\boldsymbol{\Phi}_{k}$ in (\ref{eq:o9}) can be 
 rewritten as $\boldsymbol{\Phi}_{k}=\boldsymbol{\Gamma}_{k}/(2^{r_{k}}-1)-\sum_{i=1,i\neq k}^{K}\boldsymbol{\Gamma}_{i}$,
and the objective function in (\ref{eq:bi9}) can be  re-expressed as
\begin{align}
||\mathbf{F}||_{F}^{2}+P({\bf F})=\mathrm{Tr}\left\{ \left(\mathbf{I}+{\bf D}_{F}\right)\sum_{k=1}^{K}\boldsymbol{\Gamma}_{k}\right\} +\sigma_{z}^{2}\mathrm{Tr}\left\{ \boldsymbol{\Lambda}_{\mathrm{RIS}}^{a}\right\} .
\end{align}
Since constraint $\mathrm{rank}(\boldsymbol{\Gamma}_{k})=1$ is non-convex and hard to be tackled directly. We adopt the SDR technique, that is, we first obtain an intermediate solution by dropping the rank-one constraint,  and then construct a rank-one optimal solution from the intermediate solution.   Specifically, by removing the rank-one constraints, the relaxed subproblem for  $\boldsymbol{\Gamma}=[\boldsymbol{\Gamma}_{1},\cdots,\boldsymbol{\Gamma}_{K}]$
of Problem (\ref{Pro:min-power-1})  is given by
\begin{subequations}
\label{Pro:min-power-f1}
\begin{align}
\mathop{\min}\limits _{\boldsymbol{\Gamma},\mathbf{x},\mathbf{y}} & \thinspace\thinspace\mathrm{Tr}\left\{ \left(\mathbf{I}+{\bf D}_{F}\right)\sum_{k=1}^{K}\boldsymbol{\Gamma}_{k}\right\} \\
\textrm{s.t.} & \thinspace\thinspace(\ref{eq:BTI-1}),\nonumber \\
 & \thinspace\thinspace\boldsymbol{\Gamma}_{k}\succeq\mathbf{0},\forall k, \label{42b}
\end{align}
\end{subequations}
where we have omitted all irrelative constant terms that do not depend on $\boldsymbol{\Gamma}$. Problem (\ref{Pro:min-power-f1}) is a standard SDP and can be solved using CVX. The following theorem further reveals the tightness of SDR for Problem (\ref{Pro:min-power-f1}), the proof of which can be found  in \cite[Appendix C]{GuiTSProbust}.

\begin{theorem}\label{Theorem-1} Assuming that the relaxed  Problem (\ref{Pro:min-power-f1}) is feasible,  there always
exists a feasible solution $\{\boldsymbol{\Gamma}_{k}^{\star}\}_{k=1}^{K}$
satisfying  $\mathrm{rank}(\boldsymbol{\Gamma}_{k}^{\star})=1,\forall k$.
\end{theorem}

Based on Theorem \ref{Theorem-1},  the   optimal BS beamforming vectors $\{{\bf f}_{k}^{\star}\}_{k=1}^{K}$  can be obtained from  $\{\boldsymbol{\Gamma}_{k}^{\star}\}_{k=1}^{K}$  via eigenvalue decomposition.

\subsection{Optimization of Reflection Vector ${\bf w}$ }

Next, we consider the subproblem of solving $\mathbf{w}$ for a given $\mathbf{F}$. We introduce auxiliary variable $\tilde{\mathbf{W}}=\tilde{\mathbf{w}}\tilde{\mathbf{w}}^{\mathrm{H}}$ with constraints $\tilde{\mathbf{W}}\succeq\mathbf{0}$ 	and $\mathrm{rank}(\tilde{\mathbf{W}})=1$,
thus $\boldsymbol{\Lambda}_{\mathrm{RIS}}^{a}$ defined in (\ref{eq:frfr}) and  $\mathbf{p}_{\mathrm{RIS}}$ defined in (\ref{eq:Outage-7}) can be expressed as 
 $\boldsymbol{\Lambda}_{\mathrm{RIS}}^{a}={\mathrm{Diag}}({\mathrm{diag}}([\tilde{\mathbf{W}}]_{1:M,1:M}))$
and $\mathbf{p}_{\mathrm{RIS}}=\mathrm{diag}([\tilde{\mathbf{W}}]_{1:M,1:M})$, respectively.
Correspondingly, the objective function in (\ref{eq:bf}) and the constraints in (\ref{eq:BTI-1}) are equivalent to
\begin{align}
 & P(\tilde{\mathbf{W}})=\mathbf{w}^{\mathrm{H}}{\bf D}_{w}\mathbf{w}=\mathrm{Tr}\left\{ {\bf D}_{w}{\mathrm{Diag}}({\mathrm{diag}}([\tilde{\mathbf{W}}]_{1:M,1:M}))\right\} ,\label{eq:k9}
\end{align}
and
\begin{align}
 & \left\{ \begin{array}{c}
\frac{\beta_{k}}{\delta_{k}+1}\mathrm{Tr}\left\{ {\mathrm{Diag}}({\mathrm{diag}}([\tilde{\mathbf{W}}]_{1:M,1:M}))\right\} +\sqrt{2\ln(1/\rho_{k})}x_{k}-\ln(\rho_{k})y_{k}\\
-\frac{1}{\sigma_{z}^{2}}\left(\mathrm{Tr}\left\{ \mathbf{H}_{k}\boldsymbol{\Phi}_{k}\mathbf{H}_{k}^{\mathrm{H}}\tilde{\mathbf{W}}\right\} -\sigma_{k}^{2}\right)\leq0,\forall k,\\
\frac{\beta_{k}}{\delta_{k}+1}||\mathbf{C}_{k}^{1/2}\mathrm{diag}([\tilde{\mathbf{W}}]_{1:M,1:M})||\leq x_{k},\forall k,\\
y_{k}\mathbf{I}-\frac{\beta_{k}}{\delta_{k}+1}\boldsymbol{\Sigma}_{\mathrm{r},k}{\mathrm{Diag}}({\mathrm{diag}}([\tilde{\mathbf{W}}]_{1:M,1:M}))\succeq\mathbf{0},\forall k,\\
y_{k}\geq0,\forall k.
\end{array}\right.\label{eq:BTI-3}
\end{align}
Adopting again the SDR technique
and removing non-convex constraint $\mathrm{rank}(\tilde{\mathbf{W}})=1$, we
 obtain the rank-relaxed subproblem for $\tilde{\mathbf{W}}$ of Problem
(\ref{Pro:min-power-1}) by ignoring irrelevant  constants as follows
\begin{subequations}
\label{Pro:min-power-8}
\begin{align}
\mathop{\min}\limits _{\tilde{\mathbf{W}},\mathbf{x},\mathbf{y}} & \;\;P(\tilde{\mathbf{W}})\\
\textrm{s.t.} & \;\;(\ref{eq:BTI-3}),\nonumber \\
 & \;\;1\leq[\mathrm{diag}(\tilde{\mathbf{W}})]_{m}\leq a_{max},1\leq m\leq M, \label{45b}\\
 & \;\;[\mathrm{diag}(\tilde{\mathbf{W}})]_{M+1}=1,  \label{45c}\\
 & \;\;\tilde{\mathbf{W}}\succeq\mathbf{0}, \label{45d}
\end{align}
\end{subequations}
 which is a standard SDP and can be solved  using CVX.  Since the diagonal elements of $\tilde{\mathbf{W}}$ are independently constrained in (\ref{45b}) and (\ref{45c}),  the optimal solution, $\tilde{\mathbf{W}}^{\star}$, of (\ref{Pro:min-power-8}) may not be  rank-one. Therefore, only a suboptimal $\tilde{\mathbf{w}}^{\star}$ can be constructed from $\tilde{\mathbf{W}}^{\star}$ by using the Gaussian decomposition technique. Specifically,  we consider the eigenvalue decomposition of $\tilde{\mathbf{W}}^{\star}$,  $\tilde{\mathbf{W}}^{\star}={\bf E}\boldsymbol{\Upsilon}{\bf E}^{\rm H}$, where the columns of ${\bf E}$ are the eigenvectors of  $\tilde{\mathbf{W}}^{\star}$, and diagonal matrix $\boldsymbol{\Upsilon}$ contains the corresponding eigenvalues.
 Then, we compute 1000 candidate vectors, $\{{\bf v}_{i}= {\bf E}\boldsymbol{\Upsilon}^{1/2}\mathbf{e}_{i} / [{\bf E}\boldsymbol{\Upsilon}^{1/2}\mathbf{e}_{i}]_{M+1}\}_{i=1}^{1000}$ with $\mathbf{e}_{i}\sim\mathcal{CN}(\mathbf{0},\mathbf{I}_{M+1})$, such that each ${\bf v}_{i}$ satisfies the QoS constraints. Then, the ${\bf v}_{i}$ that minimizes the total power consumption  is selected as the optimal $\tilde{\mathbf{w}}^{\star}$.  In order to ensure convergence of the proposed AO algorithm, in each iteration, we need to find a $\tilde{\mathbf{w}}^{\star}$ that decreases the objective function value compared with the previous iteration, see Section \ref{Convergence Analysis}.
 This can always be achieved empirically by generating a sufficient number of trial vectors  for Gaussian randomization.

 \subsection{Computational Complexity}
As CVX employs the interior point method,
 the  computational complexity  of solving Problems (\ref{Pro:min-power-f1}) and (\ref{Pro:min-power-8})  is  given by \cite{Ben-Tal2001convex} 
 \[
 \mathcal{O}((\sum_{j=1}^{J}c_{j}+2I)^{1/2}n(n^{2}+\underbrace{n\sum_{j=1}^{J}c_{j}^{2}+\sum_{j=1}^{J}c_{j}^{3}}_{{\rm {\mathsf{\mathrm{due\thinspace\thinspace to\thinspace\thinspace LMI}}}}}+\underbrace{n\sum_{i=1}^{I}v_{i}^{2}}_{{\rm {due\thinspace\thinspace to\thinspace\thinspace SOC}}})),
 \]
 where $n$ is the number of variables, $J$ is the number of linear
 matrix inequalities (LMIs) 
 of size $c_{j}$, and $I$ is the number of  second-order cone (SOC) constraints of size $v_{i}$. For Problem  	(\ref{Pro:min-power-f1}),  the number of variables is $n_{1}=NK$,  (\ref{eq:BTI-1}) only contains linear contraints, and (\ref{42b}) contains $K$ LMIs of size $N$. Therefore,  the approximate complexity of Problem 
(\ref{Pro:min-power-f1})  is $o_{\mathbf{F}}=\mathcal{O}([KN]^{1/2}n_{1}[n_{1}^{2}+n_{1}KN^{2}+KN^{3}])$. For Problem (\ref{Pro:min-power-8}), there are $n_{2}=M$ variables, $K$ LMIs of size $M$ and $K$ SOC of size $M$ in (\ref{eq:BTI-3}), and one LMI of size $M+1$ in (\ref{45d}).  The remaining constraints are linear. Therefore, the approximate complexity of Problem (\ref{Pro:min-power-8}) is $o_{\mathbf{e}}=\mathcal{O}([KM+2K]^{1/2}n_{2}[n_{2}^{2}+2n_{2}KM^{2}+KM^{3}])$ by neglecting terms with low-order complexity. Finally, the approximate complexity  per iteration is $o_{\mathbf{F}}+o_{\mathbf{e}}$.

\vspace{-5mm}
 \subsection{Convergence Analysis}
 \label{Convergence Analysis}
 Finally, we analyze the convergence behavior  obtained by
 alternately solving Problems (\ref{Pro:min-power-f1}) and (\ref{Pro:min-power-8}) for  solving Problem (\ref{Pro:min-power-1}).  Let $g(\bf F, \bf w)$ denote the objective value of Problem (\ref{Pro:min-power-1}). Given $\mathbf{w}^{n}$ in the $n^{\mathrm{th}}$ iteration, we have  
 	\[
 	g(\mathbf{F}^{n},\mathbf{w}^{n})\geq g(\mathbf{F}^{n+1},\mathbf{w}^{n}),
 	\]
as we can find the global optimal solution for Problem (\ref{Pro:min-power-f1}) based on Theorem \ref{Theorem-1}. Then, given $\mathbf{F}^{n+1}$, we can always find a $\mathbf{w}^{n+1}$ for Problem (\ref{Pro:min-power-8}) satisfying  	
 \[
 g(\mathbf{F}^{n+1},\mathbf{w}^{n})\geq g(\mathbf{F}^{n+1},\mathbf{w}^{n+1})
 \]
 by using the Gaussian decomposition technique. Therefore, the sequence of objective values \\ $\{g(\mathbf{F}^{n+1},\mathbf{w}^{n+1})\}$ generated in an alternating manner is monotonically non-increasing. Thus, the obtained solutions are stationary points of Problem (\ref{Pro:min-power-1}).

\section{Numerical results}

\label{simulation}

\begin{figure}
\centering \includegraphics[width=3in,height=1in]{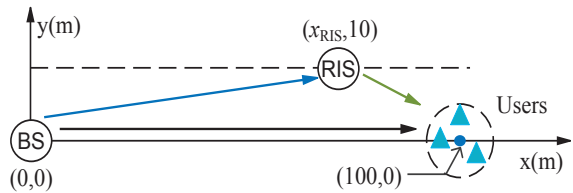}
\caption{The simulation system setup.}
\label{simulated-model}
\end{figure}

In this section, we provide numerical results to evaluate the performance
of an active RIS-aided system, where the BS and an active or passive RIS are located at (0 m, 0 m) and ($x_{\mathrm{RIS}}$ m, 10 m), as shown in Fig. \ref{simulated-model}. $K$ users are randomly and uniformly distributed in a circle with a radius of 5 m and centered at 
(100 m, 0 m). The large-scale pathloss coefficients are modelled as
$\beta=-\mathrm{PL}_{0}-10\alpha\log_{10}(d)$ dB, where $d$ is the link distance
in meters and $\alpha$ is the pathloss exponent which is set to 3.5 and 2 for the BS-user and the RIS-aided links, respectively.  $\mathrm{PL}_{0}=40$ dB denotes the pathloss at a distance of 1 meter,  i.e., we assume a carrier frequency of 3.5 GHz \cite{3Gpp-channel}.  Unless specified otherwise, the BS and the RIS are  equipped with $N=8$ antennas and $M=32$ reflecting elements, respectively, the maximum amplification gain of the  RIS is assumed to be $a_{max}=40$ dB, the Rician factors  are  $\delta_{0}=...=\delta_{K}=\delta=10$, and the noise power at the RIS and the users are set to $\sigma_{z}^{2}=\sigma_{1}^{2}=...=\sigma_{K}^{2}=-80$ dBm.

Compared with passive RISs, the power comsumed by active RISs also includes the transmit power and the circuit power for amplification. Thus,  the maximum total RIS power consumption is given by $P_{\mathrm{RIS}}=P_{M}+P_{{\rm cir}}$. Here, the circuit power $P_{{\rm cir}}=M(P_{\mathrm{c}}+P_{\mathrm{DC}})$ comprises the power  consumed by the 	phase shifters and the control circuits of the  RIS elements, $P_{\mathrm{c}}$,  and the DC biasing power, $P_{\mathrm{DC}}$, used to drive the amplifies of the
 active RIS elements. For consistency, the circuit power of each RF chain, $P_{\mathrm{RF}}$, is also accounted for in the maximum BS power comsumption, denoted by  $P_{\mathrm{BS}}=P_{N}+NP_{\mathrm{RF}}$. According to \cite{liangchang2021}, we set  $P_{\mathrm{DC}}=-5$ dBm,  $P_{\mathrm{c}}=-10$ dBm, and $P_{\mathrm{RF}}=23$ dBm.

\vspace{-5mm}
\subsection{Maximum  Achievable Rate}

\label{simulation-rate} 
\begin{figure}
	\centering \subfigure[Iteration number]{\includegraphics[width=3.4in,height=2.6in]{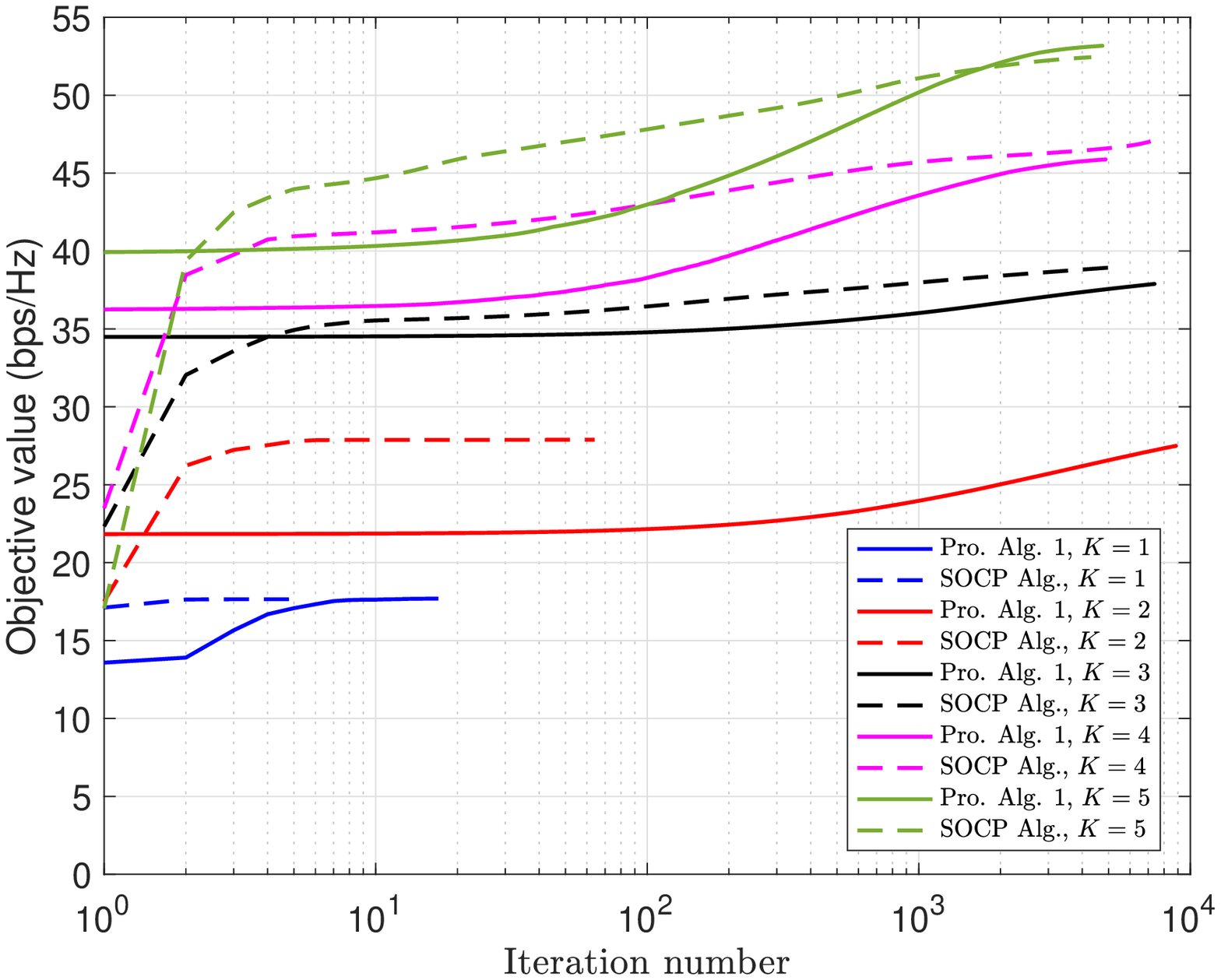}}
	\centering \subfigure[CPU time ]{\includegraphics[width=3.4in,height=2.6in]{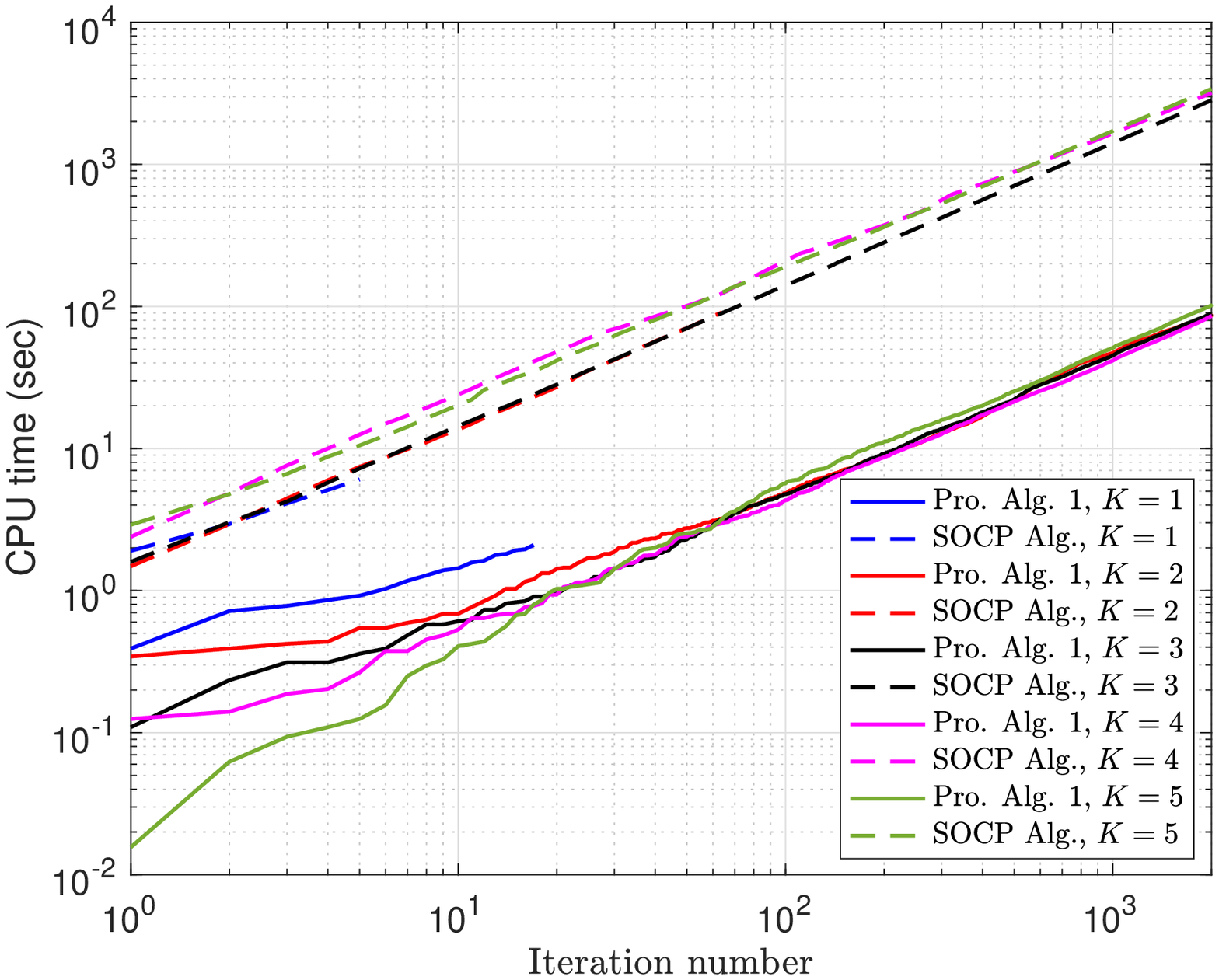}}
	\caption{ \fontsize{10.5bp}{17bp}The convergence behaviour of different algorithms for 1 randomly generated channel realization,  when $N=8$, $M=32$, and $a_{max}=40$ dB.}
	\label{objective_vs_iteration}
\end{figure}

In this subsection, we  evaluate the  ergodic achievable rate
of the active RIS-aided system as discussed in Section
\ref{average-rate} for 500 independent  realizations of $\{{{\bf h}_{k},{\bf G}_k}\}_{k=1}^{K}$, except for the convergence analysis in Fig. \ref{objective_vs_iteration}.  We denote the ergodic  achievable rate obtained by averaging the   lower bound on the average achievable rate  in (\ref{eq:rate-a}) {\footnote{  The ergodic  achievable rate and the average achievable rate should not be confused: the latter is given by $\mathbb{E}_{\mathbf{h}_{\mathrm{r},k}|{\bf G}_k}\left\{ R_{k}({\bf F},{\bf w})\right\}$ in (\ref{eq:rate-a}), while the former is given by $\mathbb{E}_{{\bf h}_{k},{\bf G}_k}\left\{ R_{k}({\bf F},{\bf w})\right\} = \mathbb{E}_{{\bf h}_{k},{\bf G}_k}\{\mathbb{E}_{\mathbf{h}_{\mathrm{r},k}|{\bf G}_k}\left\{ R_{k}({\bf F},{\bf w})\right\}\}$.  }} and the  corresponding instantaneous achievable rate in (\ref{eq:rate-1})    over all channel realizations as ``Act. RIS-LB'' and ``Act. RIS'', respectively.
For comparison,  we also consider an upper bound for ``Act. RIS'', denoted as ``Act. RIS-perfect'',  for which we assume the availability of  perfect CSI,  and a  corresponding lower bound, denoted as ``Act. RIS-non-robust'',  for which we ignore  the NLoS components of the RIS-aided channels in (\ref{eq:rate-1}) for beamformer design. In addition,  systems with passive RIS and without RIS are also considered as performance benchmarks, and are denoted as ``Pas. RIS'' and ``No RIS'', respectively.  We determine the total power consumption of the active RIS-aided system as $P_{\mathrm{BS}}+P_{\mathrm{RIS}}$, that of the
passive RIS-aided system as $P_{\mathrm{BS}}+MP_{\mathrm{c}}$, and that  of the system without RIS  as $P_{\mathrm{BS}}$. For a fair comparison, the
maximum total power consumption is set to the same value for all considered schemes.

Fig. \ref{objective_vs_iteration} {illustrates} the convergence and complexity
of the proposed Algorithm \ref{Algorithm-ao}, denoted as ``Pro. Alg. 1'', where the RIS  is located at (80 m, 10 m).  An SOCP-based algorithm  is considered as a benchmark algorithm, denoted as ``SOCP Alg.''.   For the  ``SOCP Alg.'', auxiliary variables are introduced to transfer the non-concave rate expression in the objective function to the constraints, and then SCA is used to handle the non-convex constraints. The resulting SOCP problem with multiple constraints can be   directly solved using CVX.  An SOCP based algorithms can handle optimization problems with multiple and complex constraints, but for a large number of variables, their complexity becomes high.
 As can be seen in Fig. \ref{objective_vs_iteration}(a), ``SOCP Alg.''  converges faster than ``Pro. Alg. 1'' if the number of users is small ($K=1,2$), while it loses its advantage for large numbers of users ($K=3,4,5$). Fig. \ref{objective_vs_iteration}(b) shows that the CPU time consumed by ``Pro. Alg. 1'' is significantly smaller than that of ``SOCP Alg.'', especially when the number of users is large. This is because  the number of variables in multi-user systems is high, which  causes a high computational complexity per interation in ``SOCP Alg.'', while the complexity per iteration of Algorithm  \ref{Algorithm-ao}, benefiting from semi-closed-form solutions, is low and not sensitive to the number of variables. 

\begin{figure}[htbp]
	\centering
	\begin{minipage}[b]{0.48\textwidth}
		\centering
		\hspace*{-5mm}\includegraphics[width=3.4in,height=2.6in]{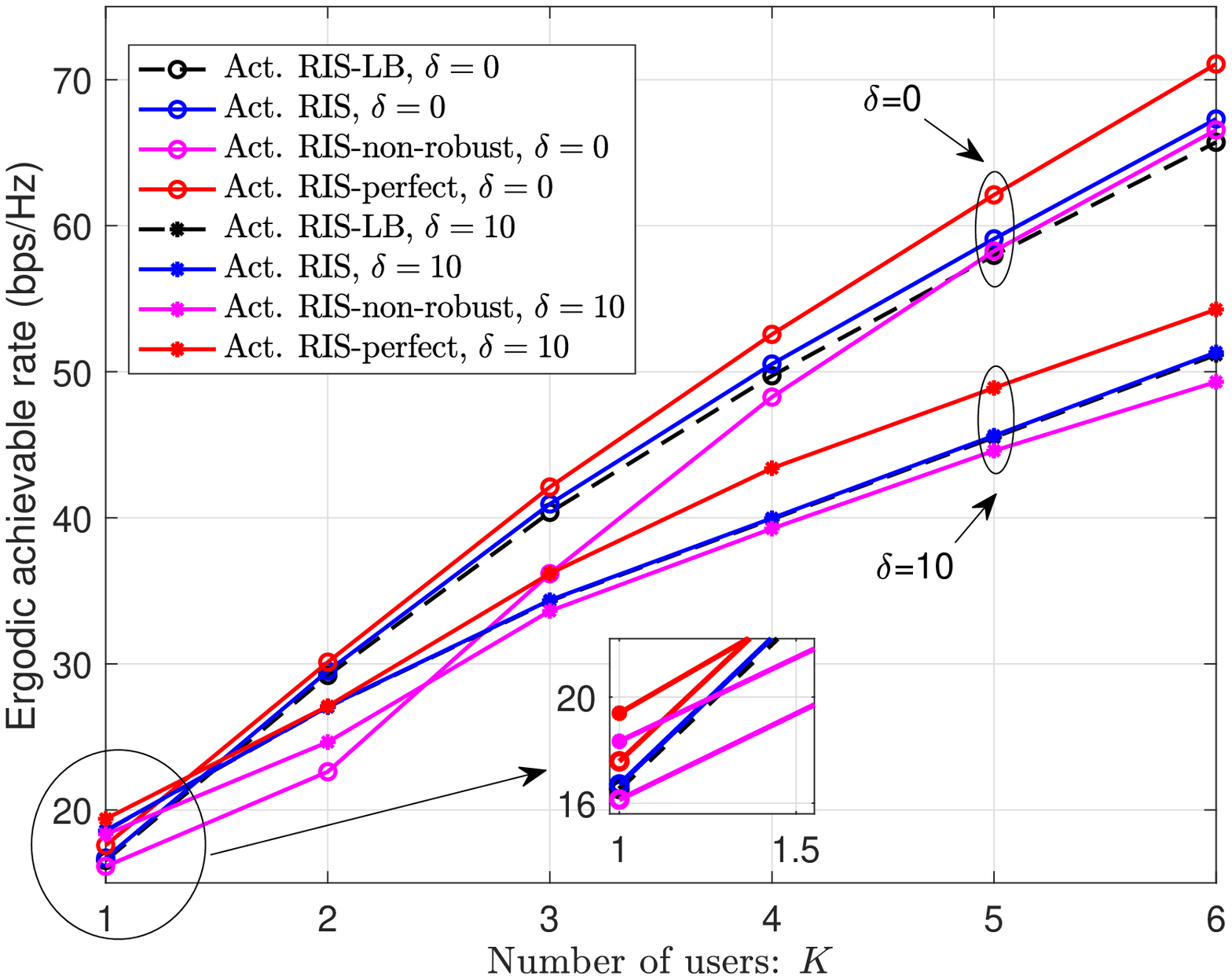}
		\caption{Achievable rate versus number of users, when $N=8$, $M=32$, and
			$a_{max}=40$ dB.}
		\label{rate-vs-user}
	\end{minipage}\hspace*{8mm}
	\begin{minipage}[b]{0.48\textwidth}
		\centering
		\hspace*{-5mm}\includegraphics[width=3.4in,height=2.6in]{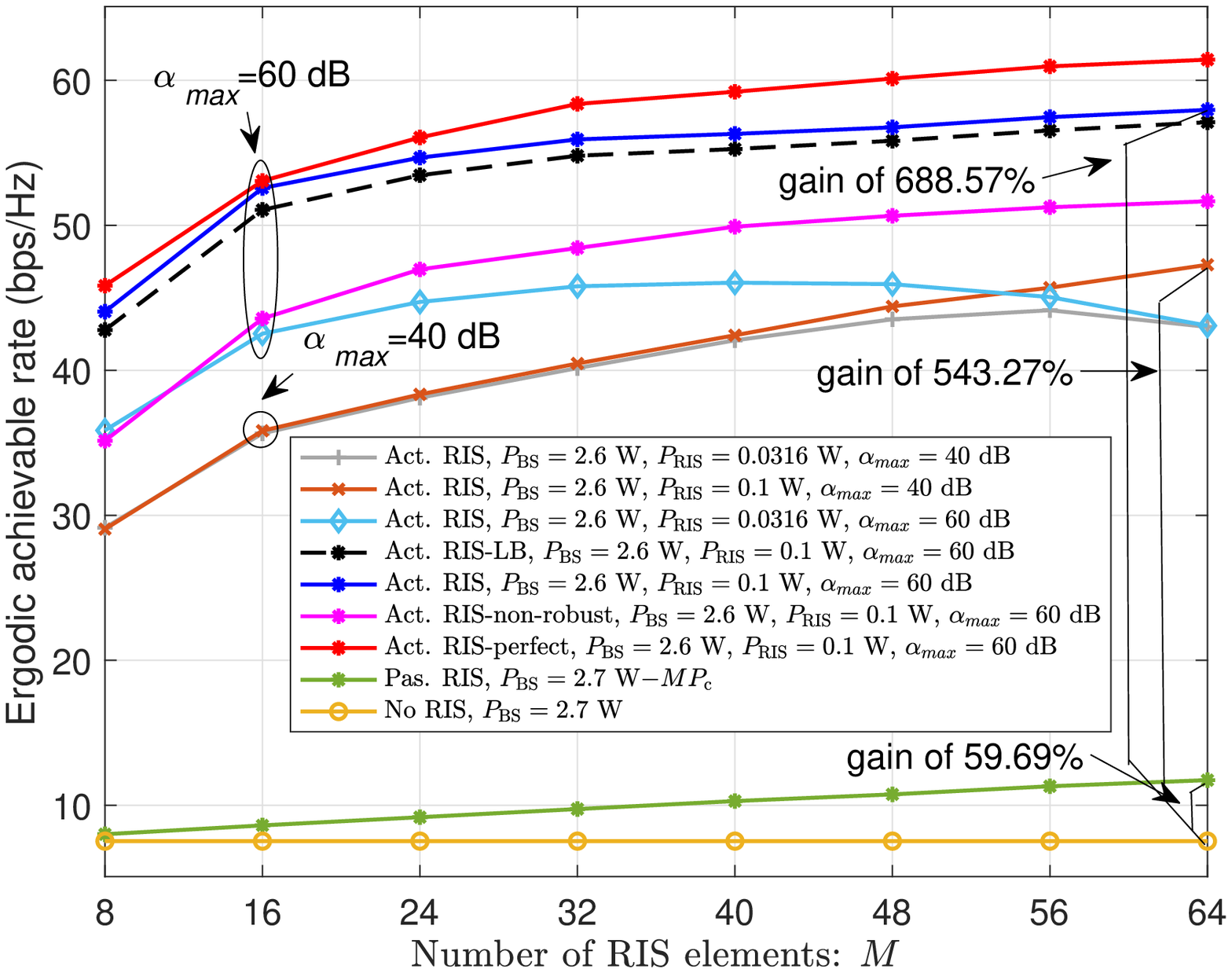}
		\caption{Achievable rate versus the number of RIS elements, when $N=8$ and $K=4$.}
		\label{rate-vs-M}
	\end{minipage}
\end{figure}



Fig. \ref{rate-vs-user} investigates the ergodic achievable rate for RIS-aided systems as a function of the number of users.
Here,  $P_{\mathrm{BS}}$ and $P_{\mathrm{RIS}}$ are set to 2.6 W and 0.1 W, respectively.
\textit{First}, as can be observed, the ergodic achievable rate  of ``Act. RIS-LB''  is only slightly lower than that of ``Act. RIS'',  when the RIS-aided channels follow a Rayleigh distribution ($\delta=0$). When the Rician factor increases to 10,  which means a reduction of the uncertain NLoS components, the ergodic achievable rate  of ``Act. RIS-LB''  is almost equal to that of ``Act. RIS''.
This confirms the tightness of the proposed lower bound expression in (\ref{eq:rate-a}).
 \textit{Furthermore}, the ergodic achievable rates for ``Act. RIS'' and ``Act. RIS-perfect''  are almost the same in the single-user case, and the gap between them increases with the number of users. This is because the negative impact of partial CSI  becomes  more significant as the number of users increases.
\textit{Finally}, the proposed ``Act. RIS'' always outperforms ``Act. RIS-non-robust'',  which reveals that a robust design is needed and that the proposed problem formulation and the corresponding algorithm  can efficiently mitigate the performance loss caused by   partial CSI.

The ergodic achievable rate as a function of the number of RIS elements
is shown in Fig. \ref{rate-vs-M} for  $K=4$ users and  with a passive or active  
RIS  fixed at (80 m, 10 m).   It is observed that the  improvement in ergodic achievable rate provided by the active RIS is affected by the amplification gain $\alpha_{max}$ and the RIS power consumption $P_{\rm RIS}$. 
\textit{First}, for $a_{max}=40$ dB, increasing  $P_{\mathrm{RIS}}$ from 0.0316 W to 1 W yields little performance improvement, which means that each reflecting element is operating with the maximum amplification gain in this scenario, i.e.,  $a_{max}=40$ dB limits the RIS power consumption even for $P_{\mathrm{RIS}}=0.0316$ W.  For $a_{max}=60$ dB, increasing  $P_{\mathrm{RIS}}$ to 1 W yields a further improvement in the ergodic   achievable rate.
\textit{Next}, for $P_{\mathrm{RIS}}=0.0316$ W, more RIS reflecting elements may actualy reduce the ergodic achievable rate.  This is because the RIS circuit power consumption increases with the number of RIS elements, and  as a result, the available RIS transmit power decreases and leads to a performance loss for $P_{\mathrm{RIS}}=0.0316$ W. Although the available RIS transmit power  is also reduced for  $P_{\mathrm{RIS}}=1$ W, the remaining transmit power is sufficient to support the additional RIS elements to achieve a performance improvement due to the resulting increased beamforming gain.
\textit{Furthermore},  for $P_{\mathrm{RIS}}=0.1$ W,  the performance gap between ``Act. RIS'' and ``Act. RIS-perfect'' increases with the number of RIS reflecting elements, which reveals a higher performance loss for RIS-aided channels with more coefficients due to the higher impact of the imperfection caused by partial CSI.  Nevertheless, compared to ``Act. RIS-non-robust'', the proposed ``Act. RIS'' can still efficiently mitigate the uncertainty of the partical CSI.
\textit{Finally},  for $M=64$ and a total power consumption of 2.7 W,  compared with the No RIS scenario, the passive RIS yields a maximum performance gain of 
59.96\%, while the active RIS with $a_{max}=40$ dB and $a_{max}=60$
dB achieves performance gains of 543.27\% and 688.57\%,
respectively. 

\vspace{-5mm}
\subsection{Minimum Average Power }

The  minimum average power consumption investigated in Section \ref{average-power} is evaluated in this subsection.  Each point in the following figures  is obtained by averaging over 500 independent channel realizations.
The maximum outage probabilities and target rates of all users are respectively assumed to be identical, i.e.,  $\rho_{1}=...=\rho_{K}=\rho=0.05$ and $r_{1}=...=r_{K}=r$.
For a fair comparison,  the total achievable power consumption including the total transmit power and the circuit power consumption, i.e., $||\mathbf{F}||_{F}^{2}+P({\bf F},{\bf w})+M(P_{\mathrm{c}}+P_{\mathrm{DC}})+NP_{\mathrm{RF}}$, is adopted as performance metric.  For the benchmark ``Act. RIS-non-robust'', the beamformer in Problem (\ref{Pro:min-power-1}) is obtained by ignoring the NLoS components of the RIS-aided channels.  

\begin{figure}
	\centering \subfigure[Total power consumption]{\includegraphics[width=3.4in,height=2.6in]{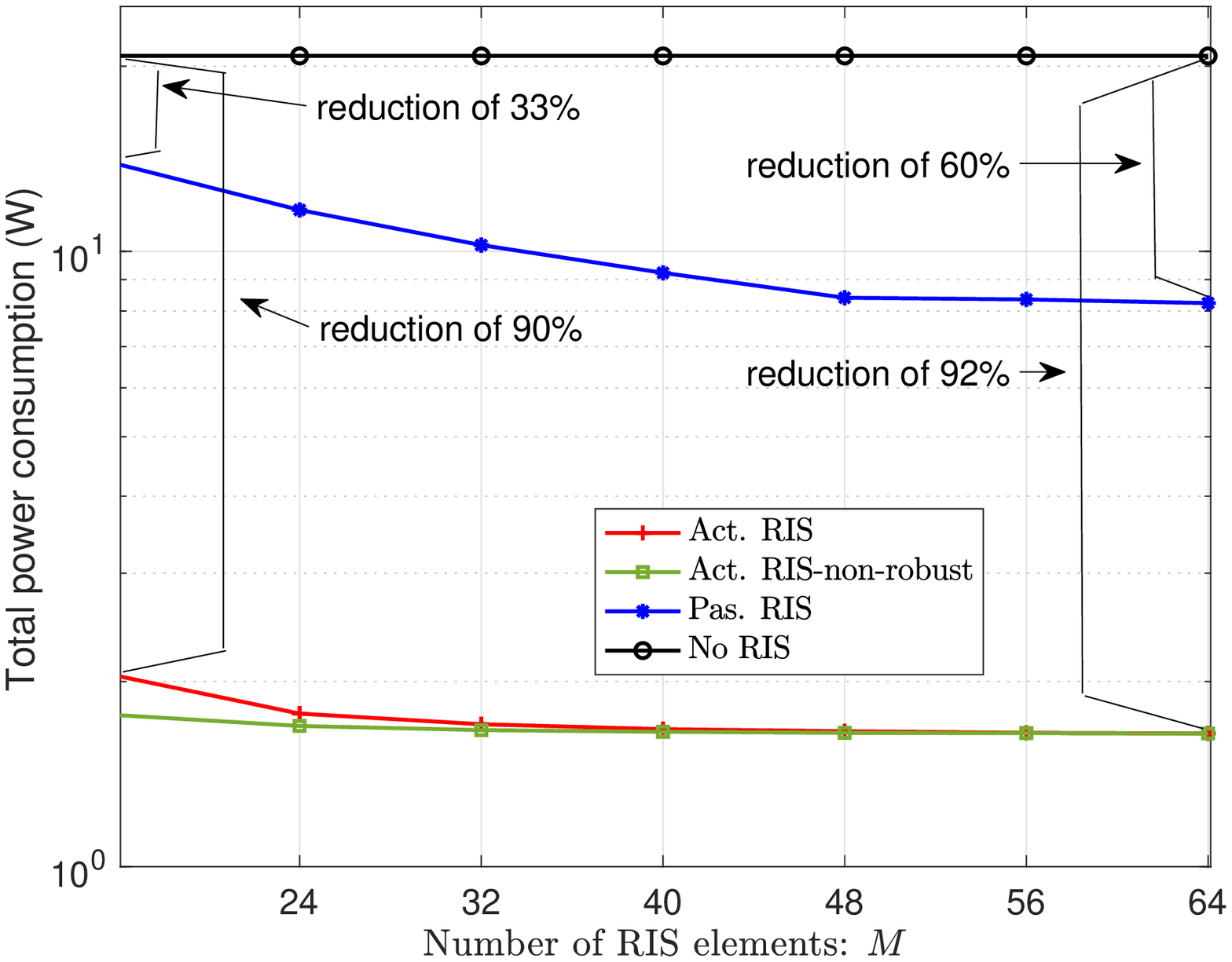}}
	\centering \subfigure[Outage probability ]{\includegraphics[width=3.4in,height=2.6in]{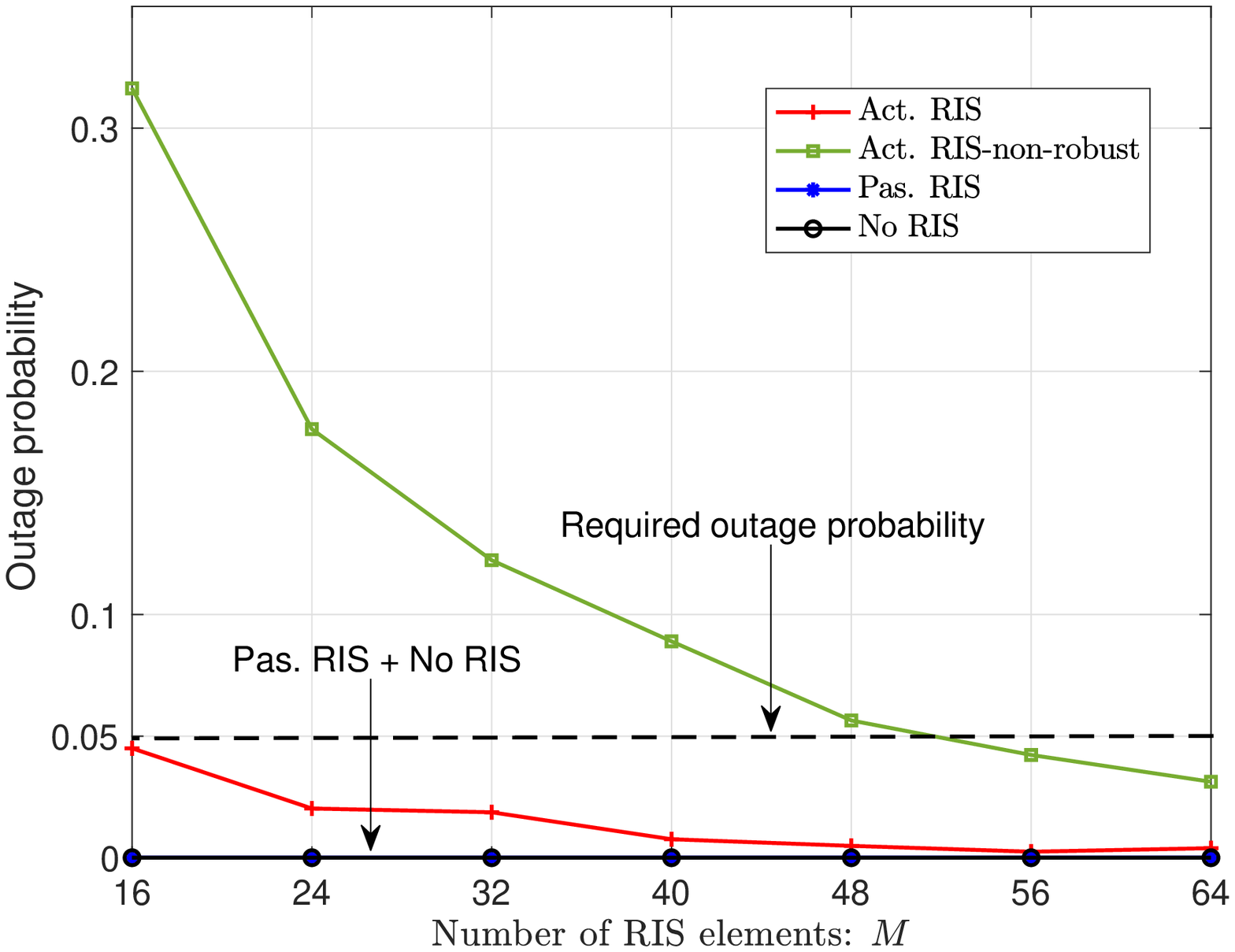}}
	\caption{ Total power consumption and outage probability versus the number of RIS elements, when $N=8$, $K=4$, $\rho=0.95$, and $r=5$ bps/Hz.}
	\label{power-vs-M}
\end{figure}

Fig. \ref{power-vs-M} shows the minimum total power consumption and the outage probability versus the number of RIS elements for a multi-user system ($K=4$), wherein the  target rate of each user is $r=5$ bps/Hz. The other parameters are set to the same values as for Fig. \ref{rate-vs-M}.
\textit{First}, as can be observed in Fig. \ref{power-vs-M}(a), an RIS equipped with only 16 active elements can reduce
the total power consumption by 90\% compared to the ``No RIS'' scenario,
while an RIS with 16 passive elements can reduce  the total power
consumption by only 33\%. Increasing the number of active reflecting elements further to $M=64$ can reduce the  total power consumption by 92\%, compared to the case  without RIS. 
\textit{Second}, the ``Act. RIS-non-robust'' scheme  consumes the least power as  the NLoS components of RIS-user links are  ignored for beamformer design. However, 	this comes at the expense of a high outage probability, cf. Fig. \ref{power-vs-M}(b).

To further demonstrate the effectiveness of the proposed ``Act. RIS'', Fig. \ref{power-vs-M}(b)   compares the outage probabilites of  ``Act. RIS'' and ``Act. RIS-non-robust''.  In particular,  the outage probability for each channel realization $\{{{\bf h}_{k},{\bf G}_k}\}_{k=1}^{K}$ is calculated as follows: For a given channel realization $\{{{\bf h}_{k},{\bf G}_k}\}_{k=1}^{K}$, 1000 conditional channel realizations $\{(\mathbf{H_{\mathrm{dr}}}, \{\mathbf{h}_{\mathrm{r},k}\}_{k=1}^{K})^{(z)}\}_{z=1}^{1000}$ are drawn from their distributions.  Then, 	the outage probability is defined as the ratio of the number of outage conditional channel realizations to the total number of conditional channel realizations, where  an outage is declared when  the target rate of at least one user cannot be satisfied. Fig. \ref{power-vs-M}(b) reveals that system outages occur frequently with  ``Act. RIS-non-robust'', especially for  low values of $M$ ($M\leq48$) due to low spatial diversity gain.   However, the proposed  ``Act. RIS''  scheme can effectively control the system outage probability to a very low level and meets the required outage probability, i.e., $\rho=0.05$, for $M\geq16$.  This illustrates the ability of the proposed scheme to mitigate the uncertainties of partial CSI. 

\begin{figure}
	\centering 
	\subfigure[Total power consumption]
	{\includegraphics[width=3.4in,height=2.6in]{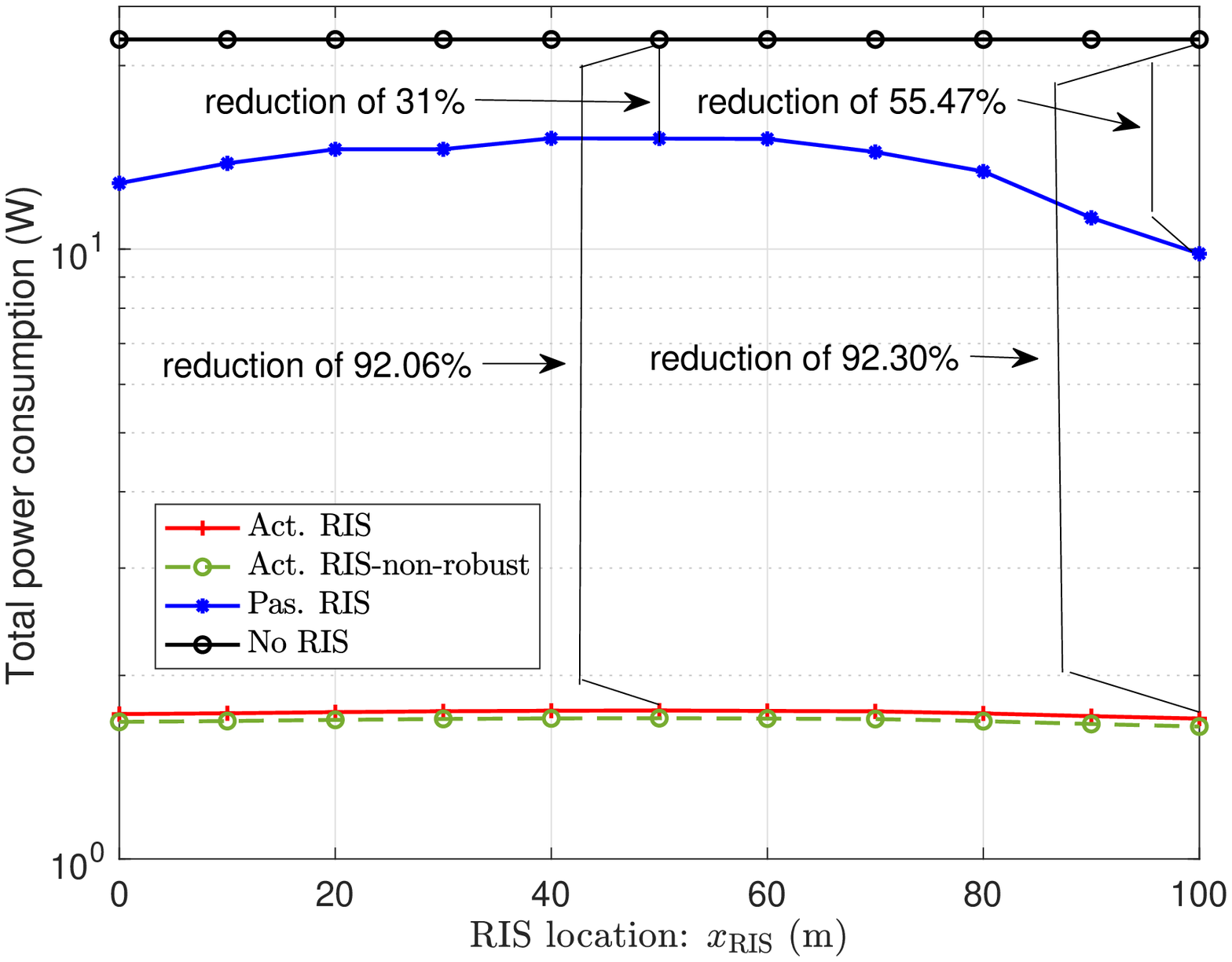}}
	\centering 
	\subfigure[Outage probability]
	{\includegraphics[width=3.4in,height=2.6in]{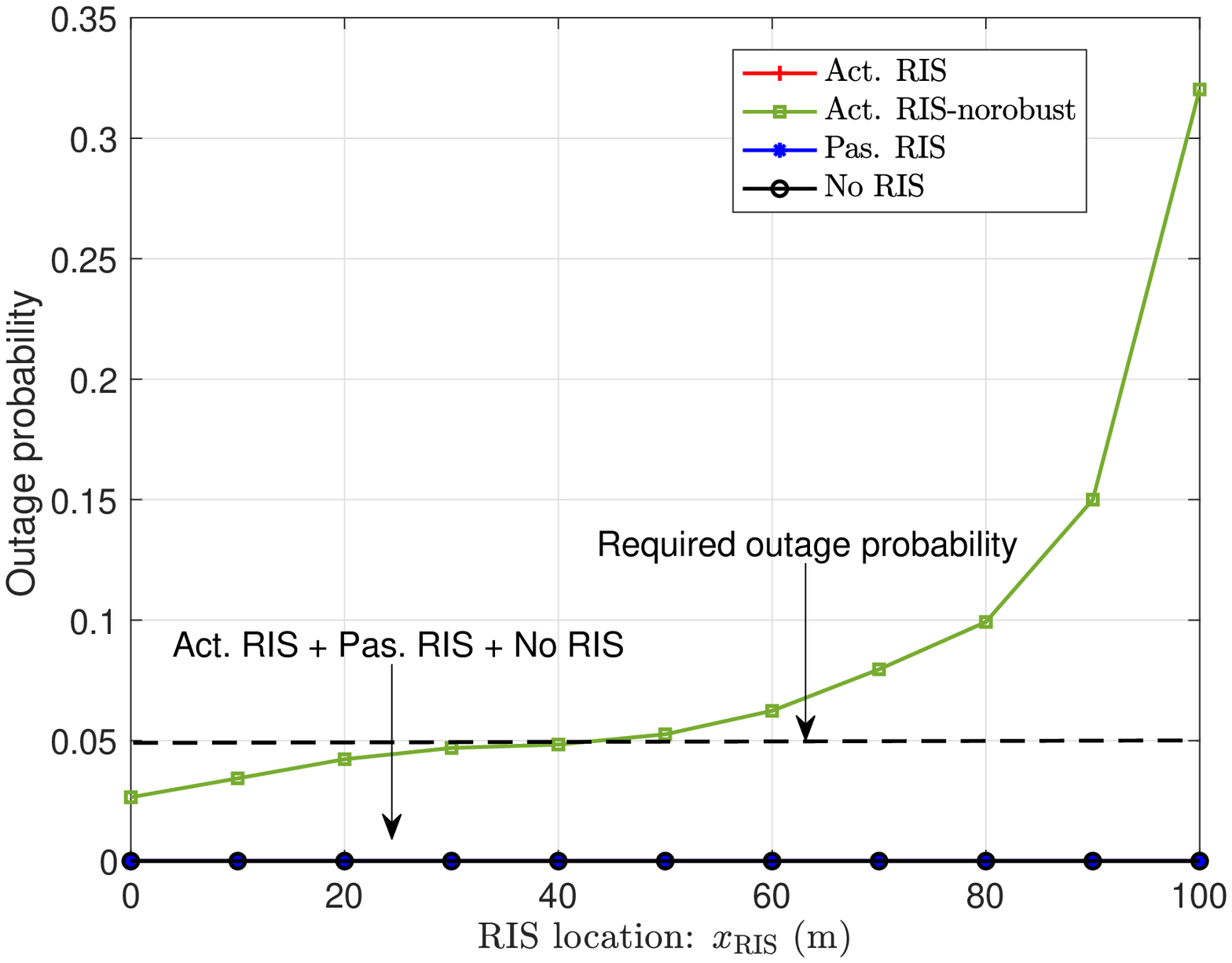}}
	\caption{Average power consumption versus RIS location, when $N=8$, $M=8$, $K=1$, $\rho=0.95$, and 	$r=10$ bps/Hz.}
	\label{power-vs-distance}
\end{figure}

To investigate the impact of the  location of the RIS on the  total power consumption and the outage probability,  Fig. \ref{power-vs-distance} considers only a single user  located at (100 m, 0 m), cf. Fig. \ref{simulated-model}.   \textit{First}, compared with the ``No RIS'' scenario, the ``Pas. RIS'' system can reduce the total power consumption by $31\%\sim55.47\%$, while the total power consumption reduction in the ``Act. RIS'' system is  considerably higher and reaches $92\%$, as shown in Fig. \ref{power-vs-distance}(a).  \textit{Second}, the ``Pas. RIS'' system has the worst performance if it is placed in the middle of the BS-user link due to the severe multiplicative fading. On the other hand, ``Act. RIS'' can siginificantly mitigate the impact of multiplicative fading in the middle of the BS-user link and yields a better performance.   \textit{Finally},   the  total power consumption of the proposed ``Act. RIS'' scheme and  ``Act. RIS-non-robust'' are almost the same, cf. Fig. \ref{power-vs-distance}(a), but, ``Act. RIS''  yields zero
 outage probability, cf. Fig. \ref{power-vs-distance}(b),  which again underscores the  benefit of the proposed scheme in mitigating the uncertainties introduced by CSI.

\section{Conclusions}

In this work, we have addressed the practical problem that perfect  individual CSI knowledge of the RIS-aided channels in active RIS systems is not available. Considering this limitation, we have derived  analytical expressions for the average achievable rate and the average RIS transmit power taking into account  the partial CSI knowledge of the individual RIS-aided channels. To  address the uncertainty caused by the partial CSI, we formulated joint BS and RIS beamforming optimization problems to respectively maximize the average sum achievable rate 
and minimize the average total transmit power subject to rate outage probability constraints.  For the average sum achievable rate maximization problem,
a computationally efficient  AO  algorithm  exploiting closed-form expressions in every iteration  has been proposed under the MM framework.  Furthermore, to facilitate the beamforming design for average transmit power minimization, we adopted the BTI to bound the rate outage probability constraints.    Subsequently, an  AO  algorithm with guaranteed convergence was developed by exploiting SDR. 
Our simulation results  confirmed that the proposed design for  average achievable rate maximization  closely approaches the performance obtained for perfect CSI. Moreover, our results revealed that compared to the non-robust scheme ignoring the unknown NLoS components, the proposed rate outage constrained design  guarantees the QoS of each user.

\appendices{}

\section{The proof of Theorem \ref{pro1}\label{appenxi1}}

To prove Property \ref{pro1}, we exploit $\mathrm{vec}(\mathbf{A}\mathrm{Diag}(\mathbf{b})\mathbf{C})=(\mathbf{C}^{\mathrm{T}}\diamond\mathbf{A})\mathbf{b}$ \cite[Equ. (1.11.21)]{Xinda2017}.
Then, we have
\begin{align}
\mathrm{Tr}\left\{ \mathbf{A}\mathrm{Diag}(\mathbf{b})\mathbf{C}\mathrm{Diag}(\mathbf{b})\right\}  & =\left(\mathrm{vec}(\mathbf{C}^{\mathrm{T}}\mathrm{Diag}(\mathbf{b})\mathbf{A}^{\mathrm{T}})\right)^{\mathrm{T}}\mathrm{vec}(\mathrm{Diag}(\mathbf{b}))\nonumber \\
 & =\left((\mathbf{A}\diamond\mathbf{C}^{\mathrm{T}})\mathbf{b}\right)^{\mathrm{T}}\mathrm{vec}(\mathrm{Diag}(\mathbf{b}))\label{eq:f0}\\
 & =\mathbf{b}^{\mathrm{T}}(\mathbf{A}\diamond\mathbf{C}^{\mathrm{T}})^{\mathrm{T}}\mathrm{vec}(\mathrm{Diag}(\mathbf{b}))\nonumber \\
 & =\mathbf{b}^{\mathrm{T}}(\mathbf{A}^{\mathrm{T}}\odot\mathbf{C})\mathbf{b},\label{eq:f9}
\end{align}
where (\ref{eq:f0}) is due to property $\mathrm{vec}(\mathbf{A}\mathrm{Diag}(\mathbf{b})\mathbf{C})=(\mathbf{C}^{\mathrm{T}}\diamond\mathbf{A})\mathbf{b}$,
and (\ref{eq:f9}) is due to
\begin{align*}
(\mathbf{A}\diamond\mathbf{C}^{\mathrm{T}})^{\mathrm{T}}\mathrm{vec}(\mathrm{Diag}(\mathbf{b})) & =\left[\begin{array}{c}
(\mathbf{a}_{1}\otimes\mathbf{c}_{1})^{\mathrm{T}}\\
(\mathbf{a}_{2}\otimes\mathbf{c}_{2})^{\mathrm{T}}\\
\vdots\\
(\mathbf{a}_{N}\otimes\mathbf{c}_{N})^{\mathrm{T}}
\end{array}\right]\mathrm{vec}(\mathrm{Diag}(\mathbf{b}))\\
 & =\left[\begin{array}{c}
(\mathbf{a}_{1}\odot\mathbf{c}_{1})^{\mathrm{T}}\\
(\mathbf{a}_{2}\odot\mathbf{c}_{2})^{\mathrm{T}}\\
\vdots\\
(\mathbf{a}_{N}\odot\mathbf{c}_{N})^{\mathrm{T}}
\end{array}\right]\mathbf{b}\\
 & =(\mathbf{A}^{\mathrm{T}}\odot\mathbf{C})\mathbf{b},
\end{align*}
where $\mathbf{A}=\left[\begin{array}{cccc}
\mathbf{a}_{1} & \mathbf{a}_{2} & \cdots & \mathbf{a}_{N}\end{array}\right]$ and $\mathbf{C}^{\mathrm{T}}=\left[\begin{array}{cccc}
\mathbf{c}_{1} & \mathbf{c}_{2} & \cdots & \mathbf{c}_{N}\end{array}\right]$.

Thus, Theorem \ref{pro1} is proved.

 \bibliographystyle{IEEEtran}
\bibliography{bibfile}

\end{document}